\documentclass[
reprint,twocolumns,
superscriptaddress,
amsmath,amssymb,
aps,prx
]{revtex4-2}

\usepackage{amsfonts}
\usepackage{soul}
\usepackage{graphicx}
\usepackage{dcolumn}
\usepackage{bm}
\usepackage{hyperref}
\usepackage{physics}
\usepackage{xcolor}
\usepackage{subfigure}

\usepackage{mathrsfs}
\usepackage{mathtools}
\usepackage{comment}
\usepackage{dsfont}
\usepackage{euscript}
\usepackage{dutchcal}
\usepackage{cellspace}

\usepackage{makecell}   
\usepackage{booktabs}   
\usepackage{array}      
\usepackage{siunitx}    

\usepackage{orcidlink}

\newcommand{\rrangle}{\rangle\!\rangle}
\renewcommand{\d}{\text{d}}

\newcommand{\fab}[1]{{\color{red}\ifmmode\text{\footnotesize(FE) #1}\else\footnotesize{(FE) #1}\fi}}

\begin{document}

\title{Complexity of quantum trajectories}

\author{Luca Lumia~\orcidlink{0000-0002-7783-9184}}
\affiliation{SISSA, via Bonomea 265, 34136 Trieste, Italy}
\affiliation{University of Strasbourg and CNRS, CESQ and ISIS, 67000 Strasbourg, France}

\author{Emanuele Tirrito~\orcidlink{0000-0001-7067-1203}}
\affiliation{Laboratory of Theoretical Physics of Nanosystems (LTPN), Institute of Physics, Ecole Polytechnique Fédérale de Lausanne (EPFL), CH-1015 Lausanne, Switzerland}
\affiliation{Center for Quantum Science and Engineering, Ecole Polytechnique Fédérale de Lausanne (EPFL), CH-1015 Lausanne, Switzerland}

\author{Mario Collura~\orcidlink{0000-0003-2615-8140}}
\affiliation{SISSA, via Bonomea 265, 34136 Trieste, Italy}
\affiliation{INFN, Sezione di Trieste, Via Valerio 2, 34127 Trieste, Italy}

\author{Fabian H.L. Essler~\orcidlink{0000-0002-1127-5830}}
\affiliation{The Rudolf Peierls Centre for Theoretical Physics, Oxford University, Oxford OX1 3PU, UK}

\author{Rosario Fazio~\orcidlink{0000-0002-7793-179X}}
\affiliation{The Abdus Salam International Center for Theoretical Physics, Strada Costiera 11, 34151 Trieste, Italy}
\affiliation{Dipartimento di Fisica “E. Pancini”, Università di Napoli “Federico II”, Monte S. Angelo, I-80126 Napoli, Italy}

\date{\today}
\begin{abstract}

Open quantum systems can be described by unraveling their Lindblad master equation into ensembles of quantum trajectories. Here we investigate how the complexity of such trajectories is affected by conservation laws and other dynamical constraints of the underlying Lindblad evolution. We characterize this complexity using a data-driven approach based on the intrinsic dimension, defined as the minimal number of variables required to encode the information contained in a data set.
Applying this framework to several systems, including variants of the quantum top and of the XXZ chain, we find that the intrinsic dimension is sensitive to the structure of their dynamics. 
The Lindblad evolution in these systems is typically chaotic; in particular, we report new signatures of autonomous chaos in the quantum top. 
At specific parameter values, however, additional constraints arise: the dynamics becomes integrable, exhibits Hilbert-space fragmentation, or develops a closed BBGKY hierarchy, leading to pronounced minima in the intrinsic dimension.
Our approach results in an unsupervised probe of the complexity of dissipative quantum systems that is sensitive to chaos and ergodicity breaking phenomena beyond the initial transient regime.

\end{abstract}

\maketitle

\section{Introduction}
Understanding and quantifying the complexity of quantum many-body dynamics is essential for several reasons. At a fundamental level, complexity governs the spreading of information and the onset of thermalization in closed quantum systems, shaping our understanding of irreversibility and emergent statistical mechanics. Furthermore, complexity  determines the extent of classical simulability.
Several quantities have been introduced to characterize its many distinct facets. Quantum resources, such as entanglement, non-stabilizerness, and non-Gaussianity, are measures of the computational complexity of the underlying dynamics~\cite{Vidal2003, Amico2008, Veitch2014, Howard2017, Takagi2018, Hebenstreit2019, Dias2024}. Krylov complexity and Nielsen’s geometric complexity, capture the scrambling of quantum information, i.e. how deeply the system explores its Hilbert space, with interesting connections to gravitational properties in the context of holographic dualities~\cite{Parker_2019, Nielsen2006, Rabinovici2023, Rabinovici2025, Susskind2014, Jefferson2017}. 

%
%

These quantifiers suggest that ergodic systems are typically complex, displaying strong information scrambling and rapid growth of the associated quantum resources~\cite{Wang2004, Vidmar2017, Nahum2017, Hashimoto2023, Leone2021, Turkeshi2025, Sierant2025, Paviglianiti2025}. In contrast, ergodicity-breaking systems are governed by the existence of additional constraints on their dynamics. The prototypical example is integrability, which gives rise to an extensive set of conserved quantities with spatially local densities~\cite{korepin1997quantum,ilievski2016quasilocal}.
Other relevant phenomena include localization, which suppresses quantum correlations and leads to slow entanglement growth and frozen operator dynamics~\cite{Nandkishore_2015, Abanin_2021}, quantum many-body scars, which generate atypical trajectories with sub-thermal entanglement within an otherwise ergodic spectrum~\cite{Turner_2018, Serbyn_2021}, and Hilbert-space fragmentation, associated to disconnected dynamical sectors and thus may reduce the overall complexity~\cite{Sala_2020, Moudgalya_2022}.

%
%
Originally formulated for unitary dynamics, these concepts have recently been extended to the domain of open quantum systems~\cite{medvedyeva2016exact, ziolkowska2020yang, essler2020integrability,robertson2021exact, de2021constructing, Paszko2025, Li_2023, Marche2025, Wisniacki_2008}.
Recent experimental and theoretical advances have 
enable the controlled engineering of dissipation channels in synthetic quantum matter and enabled increasingly accurate characterizations of open-system properties through many-body techniques, particularly in systems governed by Lindblad dynamics~\cite{Isar1994, Breuer2002, Rivas2012, Fazio2025}.
%
%
%
Integrability in the context of Lindblad equations can be defined by mapping the latter to an imaginary time Schr\"odinger equation with non-Hermitian Hamiltonian. A number of notable cases have been identified ~\cite{medvedyeva2016exact, ziolkowska2020yang, prosen2008third, Naoyuki2019dissipative,Naoyuki2019dissipativespin,robertson2021exact, de2021constructing}, including examples exhibiting “operator-space fragmentation”~\cite{essler2020integrability,robertson2021exact,li2023hilbert,Paszko2025}.
A mechanism that is a priori separate from integrability and influences the structure of open-system dynamics is the decoupling of the Bogoliubov-Born-Green-Kirkwood-Yvon (BBGKY) hierarchy, which leads to a substantial simplification in the evolution of correlation functions~\cite{Eisler2011}. This property is a particular case of operator space fragmentation and has been exploited to obtain exact results in Lindblad equations that are not necessarily integrable~\cite{Zunkovic_2014, Caspar_2016, Mesterhazy_2017, Penc2025}.{One observable consequence of the BBGKY decoupling is the absence of hydrodynamic tails in the late-time behavior of certain observables due to kinematic reasons \cite{Penc2025}.}

Ergodic and integrable dynamics both give rise to characteristic signatures in the spectrum of the time-evolution operator.
For isolated quantum systems, the famous Bohigas-Giannoni-Schmidt and Berry-Tabor conjectures connect the regularity of the dynamics to the spectral statistics of the Hamiltonian: systems with a chaotic semiclassical limit are associated to eigenvalue distributions that follow random matrix theory statistics, while integrable models follow Poisson statistics \cite{Berry1977, Bohigas1984}. 
The extension of this correspondence to open quantum maps has become known as the Grobe-Haake-Sommers (GHS) conjecture, suggesting that classically chaotic maps should arise from a Lindbladian whose complex eigenvalues follow the non-Hermitian Ginibre ensembles \cite{Grobe1988, Grobe1989}. The picture turned out to be richer. Indeed, it was later found that dissipative quantum chaos, as characterized by spectral statistics (i.e. the non-Hermitian Ginibre ensemble), could appear even in the presence of classically regular behavior~\cite{Ferrari2025, Villasenor2024}. In this manuscript, we provide a new example of this phenomenon considering an autonomous quantum top, that, notably, would be integrable both classically and without dissipation~\cite{Haake1987, Ribeiro2019}.
The analysis of the spectrum of the Lindbladian was also performed in several different models without a semiclassical limit~\cite{Akemann2019, Hamazaki_2019, Hamazaki_2020, Sà2020, Sà2021}.
%
For dissipative systems, genuinely quantum dynamical distinctions between chaotic and regular motion have been elusive. This is because the density matrix ultimately relaxes to a steady state determined solely by the dominant eigenvalue of the Lindbladian, which makes it largely insensitive to the detailed structure of the spectrum.
While the Grobe–Haake–Sommers (GHS) conjecture identifies chaos through the spectral properties of the Lindbladian, such spectral signatures do not necessarily manifest in the long-time evolution of observables.
Consequently, any signatures of dissipative chaos can typically be observed only during the transient dynamics, before the system reaches its steady state \cite{Grobe1987, Yoshimura2024, Ferrari2025, Bergamasco2023, Garcia2024}.

In this work we take a different route to characterizing ergodic and non-ergodic behavior in open quantum systems: we take inspiration from classical dynamics, where (non)ergodicity is directly observable in the structure of phase space trajectories, for example through Lyapunov sensitivity or the emergence of fractal attractors in driven-dissipative models~\cite{Strogatz2015, Wiggins2003}. 
%
{In the quantum setting, this analogy is generally approximate: it can be formalized through semiclassical phase-space representations, which allow one to capture signatures reminiscent of classical chaos. However, these features may be smoothed out or altered by genuine quantum effects, small system sizes, or strong dissipation.}

A natural framework for exploring ergodicity in open quantum systems is provided by {\it quantum trajectories} (QT), which unravel the Lindblad equation into a stochastic process that recovers the density matrix upon averaging~\cite{Gardiner2004, Wiseman2009, Jacobs2014}. 
By monitoring the quantum system along a single trajectory, the evolution consists of a smooth (non-Hermitian) dynamics, punctuated by stochastic terms dependent on the type of monitoring, that account for interactions with the external environment. When a steady state is reached, the QT do not necessarily become stationary, allowing one to observe dynamical effects beyond the initial transient regime.  
Our approach is in part motivated by the fact that in certain semiclassical models, it has been observed that QT approach the attractors of the limiting classical system \cite{Zheng1995, Brun1997}, suggesting that individual trajectories can exhibit traces of quantum chaos. Our analysis goes beyond the semiclassical limit and we will show that the properties of many-body Lindblad dynamics are reflected in the structure of the associated ensembles of QT.

We quantify the complexity of trajectories in Hilbert space by means of their \emph{intrinsic dimension}. This complexity metric has its mathematical roots in classical information theory and in the theory of fractals. It is a fundamental concept in the field of manifold learning and is commonly used as a preamble to dimensional reduction~\cite{Trunk1968, Maaten2008, Camastra2016}.
Machine learning tools have emerged as powerful methods for analyzing complex, high-dimensional data sets, which are an intrinsic property of quantum many-body physics. Their effectiveness relies on the insight that high-dimensionality often arises from the representation of the data set, while correlations can be effectively described by data points embedded on a manifold with few relevant degrees of freedom, which are what the intrinsic dimension quantifies. 
The connection of intrinsic dimension with information theory is based on Kolmogorov complexity, which sets the theoretical limit on the compressibility of information \cite{Staiger1993, Li2009}. 
This notion represents a facet of complexity that is complementary to computational complexity, which, in a quantum-mechanical context, is more closely related to quantum resources or Nielsen’s complexity.
Chaotic classical dynamical systems are known to give rise to a high Kolmogorov complexity. The rate at which this complexity grows can be quantified by the Kolmogorov–Sinai entropy, as established by Brudno's theorem~\cite{Brudno1982, Zvolev1970}.
\textcolor{black}{Their orbits often explore low-dimensional structures in phase space, such as strange attractors, whose geometry can be characterized through fractal-dimension estimators~\cite{Strogatz2015}. 
While classical attractors provide a paradigmatic example, dimensionality-estimation methods can be applied more broadly to infer the intrinsic dimension of arbitrary datasets. Their performance and limitations have been investigated in the recent data-science literature, where new estimators have also been developed~\cite{Camastra2011, Campedelli_2015, Camastra2016, Facco2017, Kamkari2024}, allowing to establish the intrinsic dimension as a complexity metric in unsupervised learning applications~\cite{Facco2019, Ansuini2019, Gong2019, Allegra2020, Pope2021, Glielmo2021, Birdal2021, Zheng2023}.}
Related ideas have been successfully applied to statistical mechanics, identifying for instance phase transitions~\cite{Mendes2021, Mendes2024, Vitale2024} or revealing hidden structures in synthetic and experimental data such as snapshots of ultra-cold atoms, outputs of quantum simulators~\cite{Verdel2024} and a  quantum scars~\cite{Cao2024}.

In our approach, datasets are constructed from ensembles of QT, allowing us to effectively quantify their complexity. 
\textcolor{black}{In contrast to semiclassical phase-space descriptions, which necessarily retain only a reduced subset of features, our analysis is performed directly on the wavefunctions. The resulting datasets therefore preserve the full information content of quantum states, allowing to probe properties that are not captured in the semiclassical limit.}
We perform a numerical investigation of classes of Lindblad equations exhibiting structurally distinct types of dynamics, with particular attention to the dichotomy between dissipative integrability and chaos, and to the decoupling of correlation functions. Our main result is to show that the intrinsic dimension effectively captures the complexity of the underlying dynamics.

This manuscript is structured as follows. Before embarking on the analysis of particular models we present a summary of our main results in Section~\ref{sec: summary}.  In Section~\ref{sec: lindblad dynamics} we then introduce the Lindblad formalism and discuss spectral properties and the Lindblad analog of the BBGKY hierarchy. This serves as a point of reference for the discussion of our results. 
In Section~\ref{sec: Id of quantum trajectories} we define QT and discuss how their complexity can be quantified in terms of their intrinsic dimension.
\textcolor{black}{Finally, we investigate several models that exemplify different mechanisms of ergodicity breaking.
In Section~\ref{sec: quantum top} we consider the quantum top, an effectively single-spin model, and study its quantum trajectories, also discussing the parameter regime where it contradicts the GHS conjecture. In Section~\ref{sec: many body} we consider trajectories of many-body spin chains, in relation to integrability and BBGKY closure.} 
Our analysis is based on a specific unraveling, namely quantum jumps. The robustness of our conclusions against alternative unravelings is examined in Section~\ref{sec unraveling}. We conclude in Section~\ref{sec conclusions}, where the main findings are summarized. 
A discussion of the semiclassical limit of the quantum top and details of our numerical analysis are relegated to appendices.

\section{Model Selection Rationale and summary of results}
\label{sec: summary}

As mentioned above, our analysis of the complexity of open system dynamics is based on the intrinsic dimension of QT, a data-driven measure that captures the spreading of QT in Hilbert space. 
We have applied this method to a set of single- and many-body models,  corresponding respectively to variants of the quantum top and the spin-1/2 Heisenberg XXZ-spin chain with dissipation.
These systems were chosen because they are generically ergodic except for some fine-tuned choices of parameters, each representing a distinct mechanism of ergodicity breaking.

The quantum top is a simple system whose degrees of freedom are defined by the components of a single angular momentum operator $S_\alpha,\,\alpha=x,y,z$, which we fix in a spin-S representation. We then consider the following time-dependent Hamiltonian
\begin{equation}
\label{eq: H quantum top}
    \hat{H}(\omega_x, \omega_z, g,k) = \hat{H}_0 + \!\! \sum_{n=-\infty}^\infty \!\!  \delta(t-n\tau)\, \hat{H}_k\,,
\end{equation}
where we have defined
\begin{eqnarray}
    \label{eq: terms quantum top}
    \hat{H}_0 &=& \omega_z \hat{S}_z + \frac{g}{S} \hat{S}_z^2 + \omega_x \hat{S}_x\ ,\quad
    H_k = \frac{k}{S} \hat{S}_y^2\,.
\end{eqnarray}
Ladder operators are defined as $\hat{S}_\pm = \hat{S}_x\pm i\hat{S}_y$. We couple the spin to a dissipative environment that polarizes it in the negative-$z$ direction. {\color{black} In Section~\ref{sec: quantum top} we show that the corresponding Lindblad equation features three different regimes: (i) quantum integrable with regular classical limit; (ii) quantum chaotic with chaotic classical limit; and (iii) quantum chaotic with regular orbits in the classical limit.}

\setcellgapes{8pt} \makegapedcells
\newcolumntype{P}[1]{>{\centering\arraybackslash}p{#1}}
\newcolumntype{C}{>{\centering\arraybackslash}c}

\begin{table}[ht!]
\centering
\begin{tabular}{|P{1.8cm}|C|}
\hline
\multicolumn{2}{|c|}{\makecell{\textbf{Driven-dissipative quantum top} }}  \\ \hline \hline
\multicolumn{2}{|c|}{
\makecell{Hamiltonian (autonomous/kicked)   \, $\hat{H}(\omega_x,\omega_z,g,k)$\\ \\
Lindblad operator     \, $\hat{L}=\sqrt{\gamma}\,\hat{S}_-$}}  \\ \hline
$\omega_x=k=0$ & Quantum and classically integrable \\ \hline
$k\neq0$ &  \makecell{Quantum chaotic, classically chaotic\\ \textcolor{black}{(see Fig.~\ref{fig: complex ratios}~(b), Fig.~\ref{fig: Idft top}~(b))}}  \\ \hline
$\omega_x\neq0$ & \makecell{Quantum chaotic, classically integrable\\ \textcolor{black}{(see Fig.~\ref{fig: complex ratios} (c), Fig.~\ref{fig: Idft top} (a))}} \\ \hline
\end{tabular}
\caption{A summary of the different parameter regimes of the driven-dissipative quantum top analyzed in Section~\ref{sec: quantum top}. The Hamiltonian is defined in Eq.(\ref{eq: H quantum top}).  Depending on the choice of the couplings $\omega_x,\omega_z,k$, different dynamical regimes can be explored.}
\label{tab:top}
\end{table}

We then turn to the many-body case and investigate the effects of integrability and of the decoupling of the BBGKY hierarchy on the intrinsic dimension
by considering appropriate perturbations of the integrable XXZ model with dephasing introduced in Ref.~\cite{medvedyeva2016exact} and of the fragmented model of Ref.~\cite{essler2020integrability}. The corresponding Hamiltonian parts are particular cases of
\begin{align}
\hat{H}(J_1,J_2,\Delta) &= \sum_{j=1}^{L-1}J_1 \Bigl[\hat{\sigma}^x_j\hat{\sigma}^x_{j+1} + \hat{\sigma}^y_j\hat{\sigma}^y_{j+1}\Bigr] + \Delta\,\hat{\sigma}^z_j\hat{\sigma}^z_{j+1}\nonumber\\
    &\,+J_2\sum_{j=1}^{L-2}\Bigl( \hat{\sigma}^x_j\hat{\sigma}^x_{j+1} + \hat{\sigma}^y_j\hat{\sigma}^y_{j+1}\Bigr)\ ,
\label{Hamiltonian}
\end{align}
where $\hat{\sigma}^\alpha_j\,,\,\alpha\in\{x,y,z\}$ denote the Pauli matrices acting on the site $j\in\{1,\,...,\,N\}$. 
{\color{black} The various Lindblad operators} we consider in Section~\ref{sec: many body} are given in Table~\ref{tab:many-body}.

Our results can be summarized as follows.
\begin{itemize}
    \item  
    For the quantum top, the intrinsic dimension of QT
   discriminates between regular and chaotic regimes: trajectories in integrable settings form effectively one-dimensional manifolds, while chaos, either induced by autonomous or time-dependent perturbations, produces higher-dimensional structures analogous to chaotic attractors. 
    Notably, the intrinsic dimension agrees with the spectral signatures of chaos in the regime with a classical limit, suggesting that the random matrix theory criterion should hold even in absence of the quantum/classical correspondence. As it will be discussed in the forthcoming sections, the quantum top suffers from severe finite size corrections, due to the fact that the different blocks of the Hilbert space grow polynomially with $S$. 

\item  
    The spatially extended many-body systems in Table~\ref{tab:many-body}
    no longer give rise to one-dimensional QT. However, the intrinsic dimension generally exhibits a sharp minimum at integrable points in parameter space, confirming that integrability structurally suppresses the complexity of QT. In cases where the BBGKY hierarchy decouples, similar reductions are observed. The presence of strong symmetries must be taken carefully into account, since it can also affect the structure of QT through the mechanism of dissipative freezing~\cite{Munoz_2019, Tindall_2022}.

\item The observations summarized above  appear to be independent of the           unraveling of the Lindblad equation. More precisely, while the               quantitative value of the intrinsic dimension is unraveling-dependent,       we find that its qualitative behavior is robust, with minima                 consistently identifying {regimes with constrained \hbox{dynamics}}.
\end{itemize}


\setcellgapes{12pt} \makegapedcells

\begin{table*}[t!]
\centering
\begin{tabular}{|P{1.7cm}|P{3.7cm}|C|C|}
\hline
\multicolumn{4}{|c|}{
{\textbf{Quantum spin chains with local interactions and local coupling to the environment}} }\\ \hline \hline 
\multicolumn{4}{|l|}{\makecell{XXZ Hamiltonian with next-nearest neighbor coupling  \, \,  $\hat{H}(J_1,J_2,\Delta)$ \\ \\
Lindblad operators     \, $\hat{L}^{(0)}_j = \sqrt{\gamma_0}\hat{\sigma}^{z}_j$ \, $\hat{L}^{(1)}_j = \sqrt{\frac{\gamma_1}{2}}(\bm{\hat{\sigma}}_j\cdot\bm{\hat{\sigma}_{j+1}}+ 1)$ \, $\hat{L}^{(2)}_j = \sqrt{\gamma_2}\hat{\sigma}^+_j\hat{\sigma}^-_{j+1}$ \, $\hat{L}^{(3)}_j = \sqrt{\gamma_2}\hat{\sigma}^-_j\hat{\sigma}^+_{j+1}$ }}
\\ \hline
Model (A) & $\gamma_1 = \gamma_2 = 0$ & Integrable with decoupled BBGKY hierarchy at $J_2=\Delta =0$ & \textcolor{black}{Fig.~\ref{fig: Idft XXZ} (a), (b)} \\ \hline
Model (B) & $J_1=J_2=0, \,\gamma_0 = \gamma_2  = 0$ & BBGKY hierarchy does not decouple; Integrable for $\Delta=0$ & \textcolor{black}{Fig.~\ref{fig: Idft XXZ} (c)} \\ \hline
Model (C) & $J_2=\Delta=0\,,\,\,\,\gamma_1 = 0$ & {Dissipative freezing} & \hspace{1pt} \textcolor{black}{Fig.~\ref{fig: diss freezing}} \hspace{1pt} \\ \hline
Model (D) & $J_2=\gamma_0 = \gamma_1 = 0$ & \hspace{4pt}Non-integrable;
BBGKY hierarchy decouples for $\Delta = 0,\,\gamma_2=\gamma_3$; \hspace{4pt} & \textcolor{black}{Fig.~\ref{fig: Idft XXZ} (d)} \\ \hline
\end{tabular}
\caption{The many-body case is represented by different variants of the XXZ Hamiltonian in the presence of next-neighboring hopping and different instances of local and correlated dissipation. The four Models (A)-(D) represent different cases where ergodicity is broken for different reasons (integrability, fragmentation) or because the BBGKY hierarchy decouples. {Model D is integrable with operator-space fragmentation for $J_1=\Delta=0$}}
\label{tab:many-body}
\end{table*}


\section{Lindblad dynamics}
\label{sec: lindblad dynamics}

The state of an open quantum system is characterized by the density matrix $\rho$, and under the hypothesis of Markovian dynamics, its time evolution is described by the Lindblad equation \cite{gorini1976completely,lindblad1976generators,Breuer2002}
\begin{equation}
\label{eq: lindblad}
    \dv{\hat{\rho}}{t} = \mathcal{L}\hat\rho\,,
\end{equation}
where 
\begin{equation}
    \mathcal{L}\hat\rho = -i [\hat{H},\hat{\rho}] + \sum_\alpha  \Bigl( \hat{L}_\alpha \hat{\rho} \hat{L}_\alpha ^\dagger -\frac{1}{2}\bigl\{\hat{L}_\alpha ^\dagger \hat{L}_\alpha, \hat{\rho}\bigr\}\Bigr)\,.
\end{equation}
Here $\hat{H}$ is the Hamiltonian acting on the system degrees of freedom, while the {\color{black} Lindblad (jump)} operators $\hat{L}_\alpha$ describe the effect of the environment. They are proportional to $\sqrt{\gamma_\alpha}$, where $\gamma_\alpha\ge0$ are the corresponding dissipation rates \cite{Breuer2002}. The dynamics can be influenced by structural properties such as integrability and the decoupling of correlation functions, that we review below.

\subsection{Dissipative integrability}
\label{ssec: linbdlad integrability}

Let us consider a one-dimensional spin chain with $L$ sites and Hilbert space $\mathcal{H}\cong(\mathbb{C}^2)^{\otimes L}$ that evolves under a local Lindblad equation.
It is convenient to employ the standard “vectorization” map $|i\rangle\langle j|\to|i\rangle|j\rangle$, whose outcome is to represent the density matrix  $\hat{\rho} \in\text{End($\mathcal{H}$)}\cong\mathcal{H}^*\otimes\mathcal{H}$ as a state vector $|\rho\rrangle\in\mathcal{H}\otimes\mathcal{H}$. In this representation, super-operators are represented by matrices on $\mathcal{H}\otimes \mathcal{H}$, which we denote using a check symbol. The Lindblad equation then takes the form 
\begin{equation}
\label{eq: vectorized lindblad}
\frac{d}{dt}|\rho\rrangle = \check{\mathcal{{L}}} |\rho\rrangle \equiv (-i\check{\mathcal{H}}+\check{\mathcal{D}})|\rho\rrangle \,,
\end{equation}
where 
\begin{gather*}
    \check{\mathcal{H}}=\hat{H}\otimes\mathbb{\hat{I}} - \mathbb{\hat{I}}\otimes \hat{H}^*\,,\\
    \check{\mathcal{D}} = \sum_\alpha \Bigl(\hat{L}_\alpha\otimes \hat{L}_\alpha^* -\frac{1}{2}\bigl[\hat{L}_\alpha^\dagger \hat{L}_\alpha\otimes\mathbb{\hat{I}} + \mathbb{\hat{I}}\otimes (\hat{L}_\alpha^\dagger \hat{L}_\alpha)^*\bigr]\Bigr)\,.
\end{gather*}
The Lindbladian $\check{\mathcal{L}}$ can then be interpreted as a non-Hermitian Hamiltonian of a system with doubled degrees of freedom, in this case a 2-leg ladder \cite{ziolkowska2020yang}, and a sufficient condition for integrability is then that $\hat{\cal L}$ can be obtained from a solution of the Yang-Baxter equation \cite{korepin1997quantum}. The Lindblad structure of $\check{\cal L}$ imposes constraints in addition to integrability, but many examples have been identified in recent years~\cite{medvedyeva2016exact, ziolkowska2020yang, essler2020integrability, robertson2021exact, de2021constructing}.
A different mechanism that can significantly constrain Lindbladian dynamics is operator-space fragmentation \cite{essler2020integrability}, which is the open quantum system analog of Hilbert-space fragmentation \cite{Moudgalya_2022}. In fragmented models, there is a large number of subspaces that are left invariant by the dynamics. In some cases the projections of the Lindbladian onto these “fragments” can be integrable \cite{essler2020integrability,robertson2021exact}. 

An important difference between the quantum integrability of Schrödinger and Lindblad equations arises from their symmetries. In integrable models describing unitary time evolution it is possible to construct an extensive set of quasi-local operators that commute with the Hamiltonian and one another. 
In integrable dissipative models one instead obtains super-operators that commute with the Lindbladian and one another. This condition identifies a \emph{weak symmetry} and is no longer sufficient to define conservation laws: in this context the expectation values of operators are preserved only by \emph{strong symmetries} defined by operators that commute separately with the Hamiltonian and with all Lindblad operators~\cite{Buca2012, Albert2014}.

The distinction between integrable and chaotic behavior is typically reflected in its level-spacing distributions. 
Having complex eigenvalues, the relevant spacings are the distances $s_j = |\lambda_j - \lambda_j^{nn}|$ between each eigenvalue $\lambda_j$ and its nearest neighbor $\lambda_j^{nn}$ in the complex plane, which are distributed according to generalizations of the Poisson/Wigner-Dyson statistics of chaotic/integrable Hamiltonians  \cite{Grobe1988}. For an integrable Lindbladian, $s$ is distributed according to a 2d Poisson process in the complex plane
\begin{equation}
    p_{P}^{(2D)}(s) = \frac{\pi s}{2}\,e^{-\frac{\pi}{4}s^2}\,.
\end{equation}
Even though $p(s)\propto s$, there is no level repulsion in the complex plane because the $s$ is a Jacobian factor. 
Chaotic Lindbladians instead have spacings distributed like those of the eigenvalues of non-Hermitian random matrices. These are described by the Ginibre ensemble 
\begin{equation}
	p_{\text{GinUE}}(s) =\prod_{k=1}^{\infty} \frac{\Gamma(1+k,s^2)}{k!} \sum_{j=1}^{\infty}  \frac{2s^{2j+1} e^{-s^2}}{\Gamma(1+j,s^2)}\,, 
\end{equation}
where $\Gamma(1+k,s^2)=\int_{s^2}^{+\infty} t^k e^{-t}$ is the incomplete Gamma function. Expanding for $s\to0$ results in a $p(s)\sim s^3$ level repulsion, which is independent of the symmetry class for the three Ginibre ensembles GinOE, GinUE, GinSE, unlike for the unitary ensembles GOE, GUE, GSE~\cite{Grobe1988}.
In practice, working with a single Lindbladian, the distribution of its eigenvalues is going to be dominated by their density, which is highly model dependent, and it is the distribution of the “unfolded” eigenvalues that should follow the random matrix theory ensembles \cite{Haake2010}. Unfolding amounts to removing the effect of the local density from the spectra and is a standard procedure for real eigenvalues, but it is more delicate in the complex case. It is convenient to introduce the complex ratios
\begin{equation}
\label{eq: complex ratio}
	z_j = \frac{\lambda_j-\lambda_j^{nnn}}{\lambda_j-\lambda_j^{nn}} = r_j e^{i\theta_j},
\end{equation}
{where $\lambda_j^{nnn}$ is the next-to-near neighbor of $\lambda_j$ in the complex plane.}
{This removes the dependence on the local density} and implicitly unfolds the spectrum \cite{Sà2020}. Under the hypotheses of Poisson/Ginibre spectra, the distribution of $z_j$ visibly changes, as the level repulsion results in a suppression of the probabilities of finding $z_j$ close to the origin of the plane and $\theta$ close to 0. An example of this difference is shown in Fig.~\ref{fig: complex ratios}, where we apply this criterion to identify regular and chaotic regimes of a dissipative quantum top analyzed further in Sec.~\ref{sec: quantum top}. A simple quantitative way to quantify the distinction is to calculate the average $\overline{\cos\theta}$, equal to $0$ in the Poisson case, and $\overline{\cos\theta}\approx-0.24$ for the Ginibre ensemble.

\subsection{Lindblad BBGKY hierarchy}
\label{ssec: bbgky}
A particularly simple class of models with operator space fragmentation are Lindblad equations with decoupled BBGKY hierarchies
for the equations of motion for correlation functions. These exhibit significant simplifications in their dynamics, which are a priori not related to integrability.
Their structure is most easily explained by considering models of spinless fermions with creation annihilation operators $c^\dagger_i, c_j$ that fulfil $\{c^\dagger_i,\,c_j\}=\delta_{ij}$.
The expectation value of a generic observable $O$ evolves under the adjoint Lindbladian
\begin{equation*}
\label{eq: adjoint lindblad}
    \frac{d}{dt}\langle O\rangle = i\langle [H,O]\rangle +\sum_{\alpha} \Bigl(\langle L^\dagger_\alpha O L_\alpha \rangle - \frac12\langle\{L^\dagger_\alpha L_\alpha, O\}\rangle\Bigr)\,.     
\end{equation*}
The operators \mbox{$O^{n,m}_{\{i\},\{j\}}=c^\dagger_{i_1}\,...\,c^\dagger_{i_n}c_{j_1}\,...c_{j_m}$} form a basis and obey a hierarchy of coupled equations of motion
\begin{equation*}
    \frac{d}{dt}\langle O^{n,m}_{\{i\},\{j\}}\rangle =\sum_{\overset{r,s}{\{k\},\{l\}}}\mathcal{K}^{n,m}_{r,s}(\{i\},\{j\}|\{k\},\{l\})\,\langle O^{r,s}_{\{k\},\{l\}}\rangle\,,
\end{equation*}
\textcolor{black}{where the specific expression of the kernel $\mathcal{K}_{r,s}^{n,m}$ can in principle be derived by explicit substitution of $O^{n,m}_{\{i\},\{j\}}$ into the adjoint equation above. In general, exponentially many terms in $L$ can be generated. 
However, when the set of jump operators is closed under conjugation, {$\{L_\alpha^\dagger\}=\{L_\beta\}$}, the dissipator can be written as a double commutator} 
\begin{equation}
\begin{split}
    \frac{d}{dt}\langle O^{n,m}_{\{i\},\{j\}} \rangle_t  =&\, i\,\langle[H,O^{n,m}_{\{i\},\{j\}}]\rangle_t \\
   &-\frac{1}{2}\sum_\alpha\langle[L_\alpha^\dagger,[L_\alpha,O^{n,m}_{\{i\},\{j\}}]]\rangle_t\,.
\end{split}
\end{equation}
If we assume that both the Hamiltonian and the jump operators are quadratic in the fermionic operators, the equations above preserve the number of fermions: an operator can only mix with others carrying the same number of fermion creation/annihilation operators, resulting in a decoupling of the hierarchy of correlation functions~\cite{Eisler2011}. This structural simplicity has been used to obtain exact results in non-integrable Lindblad models~\cite{Zunkovic_2014, Caspar_2016, Mesterhazy_2017, Penc2025}, and it is natural to ask whether an appropriate measure of complexity could serve as a witness of this dynamical simplification.

\section{Quantum trajectories and their intrinsic dimension}
\label{sec: Id of quantum trajectories}

Open quantum systems can be described by unraveling their master equation into QT. This entails defining a stochastic evolution on the Hilbert space, in such a way that averaging over the realizations of the stochastic process recovers the dynamics of the density matrix. A single realization of the stochastic evolution is called a {\it quantum trajectory}. Unraveling has been proven to be a very useful numerical tool for the integration of Lindblad equations and is also experimentally relevant, since it accounts for the the dynamics of quantum systems subject to monitoring~\cite{Gardiner2004, Wiseman2009, Jacobs2014}. Physical properties of QT have recently attracted attention, fueling studies for instance on their thermodynamics \cite{Garrahan2010, Carollo2019, Perfetto2022, Pigeon2025} and on measurement-induced phase transitions \cite{Li2019_MIPT, Skinner2019, Gullans2020}. The asymptotic structure of quantum trajectories has also been recently studied \cite{Benoist_2021, Schmolke_2025}.
The aim of this paper is to investigate whether QT inherit any property from the dynamical structure of the associated Lindbladian. In particular, we will analyze their complexity as quantified by their intrinsic dimension. 
Such a correlation is not a priori expected for integrable Lindblad equations, because weak symmetries do not typically imply the existence of conservation laws that constrain QT~\cite{Albert2014,Munoz_2019}.

There is an infinite number of ways to unravel the Lindblad equation. Below we discuss the quantum jump unraveling and the corresponding trajectories. Afterwards we will explain their data set representations and introduce the intrinsic dimension as a measure to quantify their complexity.

\subsection{Quantum jumps}
\label{sec: quantum jumps}
Quantum jump unravelings arise experimentally, for example, from photon-counting processes: the system is coupled to a monitoring apparatus, and evolves deterministically until the detector measures a photon that was entangled with the system, inducing a sudden collapse of its wavefunction called quantum jump \cite{Wiseman2009}. The jump occur at random times and the dynamics is described by the stochastic process illustrated below (consider for simplicity an initial pure state $|\psi_0\rangle$). For each time increment $dt$, the wavefunction evolves as $|\psi_t\rangle\,\to\,|\psi_{t+dt}\rangle$, with
\begin{equation}
    |\psi_{t+dt}\rangle = \left\{
    \begin{tabular}{ c c c }
 $\frac{L_\alpha |\psi_t\rangle}{|\!| L_\alpha |\psi_t\rangle |\!|}$ & with probability  & $p_{\alpha}$ \\ 
 $\frac{e^{-idt\,H_{\rm{eff}}} |\psi_t\rangle}{|\!|e^{-idt\,H_{\rm{eff}}} |\psi_t\rangle|\!|}$ & with probability &  $p_0$ \;
\end{tabular} \right. \;. 
\label{qjunraveling}
\end{equation}
In Eq.(\ref{qjunraveling}), $p_\alpha =  dt\, \langle \psi_t | L_\alpha^\dagger L_\alpha | \psi_t\rangle$, $p_0 = 1 - \sum_{\alpha} p_{\alpha}$, and the non-Hermitian Hamiltonian
$
 H_{\rm{eff}} = H-\frac{i}{2}\sum_\alpha L_\alpha^\dagger L_\alpha\,
$.
%
%
The non-Hermitian term in $H_{\rm{eff}}$ ensures the conservation of probability in the evolution. Each realization of the stochastic process is characterized by the sequence ${\cal N} \equiv \{(t_1,\alpha_1); (t_2,\alpha_2); \dots , (t_n,\alpha_n); \dots \}$ with $t_i \in[0,T]$.
The mean state is defined as the average over all possible QTs $\rho_t\equiv E_\mathcal{N}[|\psi_t\rangle_{\mathcal{N\,N}}\langle\psi_t|]$. It is straightforward to show that $\rho_t$ evolves according to the Lindblad equation in  Eq.(\ref{eq: lindblad}). While averages of observables can be equivalently computed through QTs or the Lindblad evolution, non-linear functionals of the state, such as higher moments $E_\mathcal{N}[(|\psi_t\rangle_{\mathcal{N\,N}}\langle\psi_t|)^k]$, are not predicted by the Lindblad equations, indicating that QT actually encode more information than the Lindblad equation. The intrinsic dimension defined 
in the next subsection, the key quantity for our characterization of the complexity of QTs, is a non-linear function of the state.

\subsection{Intrinsic dimension}
\label{ssec: intrinsic dimension}

\begin{figure}[b]
\centering

    \includegraphics[width=0.48\textwidth]{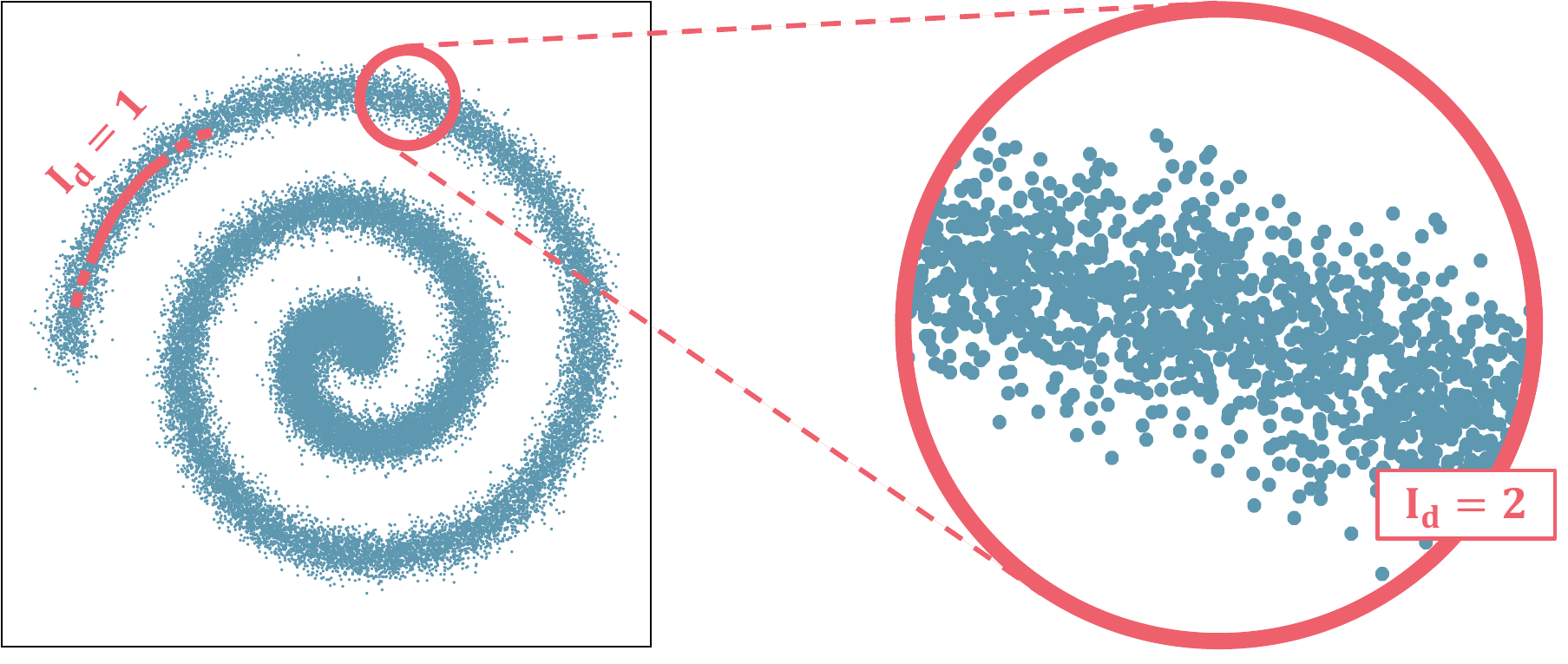}
    
\caption{Illustration of the scale dependence of the intrinsic dimension. If the nearest neighbours lie at a distance comparable to the scale of local noise, the structure appears two-dimensional. Only at an appropriately large scale the underlying curve emerges.} 

\label{fig: Id scale dependence}
\end{figure}

The intrinsic dimension $I_d$ is a quantifier of the minimal number of independent parameters that are necessary to characterize a given data set without significant loss of information. 
%
An abstract data set can be represented by a sequence of vectors $\{x_i\}_{i=1...N}$, where $x_i\in\mathbb{R}^D$ and $D$ is called the \emph{embedding dimension}.
%
The embedding dimension depends on how data points are represented: a sub-optimal encoding of the information contained in the points would result in an inefficient representation.  As a result of correlations, data points are often restricted to a submanifold of $\mathbb{R}^D$, and the {intrinsic dimension} is defined as the dimensionality of this submanifold. This is a typical property of complex data sets: an example in Physics is that while ground states of many-body quantum systems are mathematically represented as vectors in a high-dimensional Hilbert space, they often allow for compressed representations exploiting their entanglement structure.

The estimation of $I_d$ is an active field of research 
and there are many approaches to it, each with its own advantages and disadvantages \cite{Camastra2016}. The definition given above is not practical, since the embedding manifold is a priori unknown and the $I_d$ must be inferred from the correlations of a discrete set of points. 
The  correlation dimension introduced by Grassberger and Procaccia for fractals \cite{Grassberger1983} is a suitable estimator, but suffers from a density dependence problem. Other relevant approaches involve principal component analysis (PCA), multi-dimensional scaling (MDS) methods, neural networks \cite{Tenenbaun2000, Cox2000, Kirby2000, Jolliffe2016, Bahaldur2019}.
We focus on the \emph{{\tt 2-NN} method}, which estimates $I_d$ using the distances to the first and second nearest neighbors of each data point. We briefly describe the algorithm, referring the reader to the original paper for a more detailed discussion \cite{Facco2017}. Given a distance function $d$ on the embedding space, for each data point $x_i$ one calculates the distance to its nearest neighbor $r_i^{nn} = d(x_i,x_i^{nn})$ and to its next-to-nearest neighbor $r_i^{nnn} = d(x_i,x_i^{nnn})$. One then defines the ratio ${\mu_i = r_i^{nnn}/r_i^{nn}\ge 1}$, which removes the influence of different densities in distinct regions of the data set. Assuming that the neighbors reside in a small ball with dimension $I_d\le D$ where points are uniformly distributed (formally from an $I_d$-dimensional Poisson process), the ratios $\mu$ follow a Pareto distribution $p(\mu) = I_d \,\mu^{-I_d-1}$. Its cumulative distribution function $F(\bar\mu)\equiv\text{prob}\{1\le \mu \le \bar\mu\}$ satisfies
\begin{equation}
\label{eq: Id fit}
    -\log[1-F(\mu)] = I_d \log\mu\,,
\end{equation}
which implies that $I_d$ can be estimated from a simple linear regression of $\{(\log\mu_i, -\log[1-\hat{F}(\mu_i)])\}_{i=1...N}$, where $\hat{F}$ is 
the empirical cumulative distribution function. The assumption of uniformity is important, but mild, because the data set only needs to be  \emph{locally} uniform on the scale of nearest-neighbors. In practice, one can check that this hypothesis is satisfied from the accuracy of the linear fit described above: real data sets typically show some minor deviations, and the regression is then performed over the linear region \cite{Facco2017}. 

An important fact to note is that $I_d$ is scale dependent. This behavior can be easily explained by looking at toy examples. Consider a simple data set embedded in $\mathbb{R}^2$ constructed by sampling points from a 1d curve and adding some Gaussian noise, \emph{cf.} Fig.~\ref{fig: Id scale dependence}. On short scales the noise spreads the data points in the plane, resulting in a two-dimensional data set. However, on scales that are large compared to the variance of the noise the underlying curve emerges. The scale that the {\tt 2-NN} method probes can be quantified by the typical nearest-neighbor distance $\bar{r} = 1/2 \cdot \overline{r^{nn} + r^{nnn}}$, where the bar denotes the average over the data set. In the previous example $I_d=I_d(\bar{r})$ would be a decreasing function, but the opposite behavior is also possible. For example, the data set obtained by sampling points from a space filling curve such as the Hilbert curve would appear two-dimensional if the points are not too dense, while on shorter scales the underlying one-dimensional structure becomes evident. In general, studying the scale dependence of $I_d$ provides complementary information on the data manifold \cite{Facco2017, Denti2022, Recanatesi2022, Macocco2024}.

\subsection{Quantum trajectories as data sets}
\label{ssec: data sets}

We start from a fixed, randomly chosen initial pure state $|\psi_0\rangle$. Given a set of $N$ QT $|\psi_t\rangle_{\mathcal{N}_i}\,,\,i=1,\dots,N$ at least two meaningful intrinsic dimensions could be considered. 
The first is the $I_d$ of individual QT. Every trajectory is a piecewise continuous curve (with discontinuities possibly arising at jump times), therefore intrinsically one-dimensional. However, akin to what happens with classical dynamical systems and chaotic attractors, one-dimensional trajectories can be space-filling and cover a higher-dimensional portion of the Hilbert space, as we will show in Section~\ref{sec: quantum top}. In this case, the data set associated to the $i$-th trajectory is $\mathcal{D}_i \equiv \{|\psi_0\rangle,\,|\psi_{\d t} \rangle_{\mathcal{N}_i},\,|\psi_{2\d t} \rangle_{\mathcal{N}_i}, 
\dots\}$, up to a final time that is long enough to stabilize the result. 
\textcolor{black}{To perform the analysis described in Section~\ref{ssec: intrinsic dimension}, we have to store the information contained in wavefunctions --naturally living in a complex Hilbert space $\mathcal{H}$-- in real vectors.
For most of this work, we choose to do so by separating the real and imaginary parts of each of their components; as a consequence, the embedding space that we employ is $\mathbb{R}^D,\, D=2\dim{\mathcal{H}}$. Other possibilities exist: for instance one could represent the quantum state by storing expectation values of Hermitian observables. Indeed, this is what we will employ in Fig.~\ref{fig: trajectories}, where we consider individual trajectories represented only by few expectation values, in order to build an intuition.}


The rest of our analysis will be dedicated to the intrinsic dimension of the entire set of trajectories as a function of time: at a given time $t$, different QT reach different points spread out in the Hilbert space, and we estimate the dimensionality of the structure they are covering. To estimate $I_d$ at a time $t$ we consider the data set
\begin{equation}
\label{eq: data set}
    \mathcal{D}_t \equiv \{|\psi_t\rangle_{\mathcal{N}_1},\,|\psi_t\rangle_{\mathcal{N}_2},\dots,\,|\psi_t\rangle_{\mathcal{N}_N}\}
\end{equation} 
and the nearest-neighbors are classified according to the Euclidean distance of the normalized wavefunctions
\begin{equation}
    d(\psi,\phi)=\sqrt{|\!| |\psi\rangle-|\phi\rangle |\!|^2}\,.
\end{equation}
We then compute $I_d(t) = I_d[\mathcal{D}_t]$ at several times $t$ in order to observe the evolution of the complexity of QT. 
The advantages of this approach are twofold. First, it 
provides us with a single, time-dependent measure of complexity that is not associated with individual trajectories.
Second, the points used to construct the data sets are uncorrelated in time, which removes the influence of the underlying one-dimensional structure of individual trajectories. This results in a more direct applicability of the Pareto assumption.
%
A disadvantage of this approach is its higher computational cost, which arises because the calculation of $I_d$ cannot be run in parallel for every individual trajectory. 
In Appendix~\ref{app: Id single traj} we compare the results obtained from the two methods.

\begin{figure*}[]
\centering

    \includegraphics[width=0.95\textwidth]{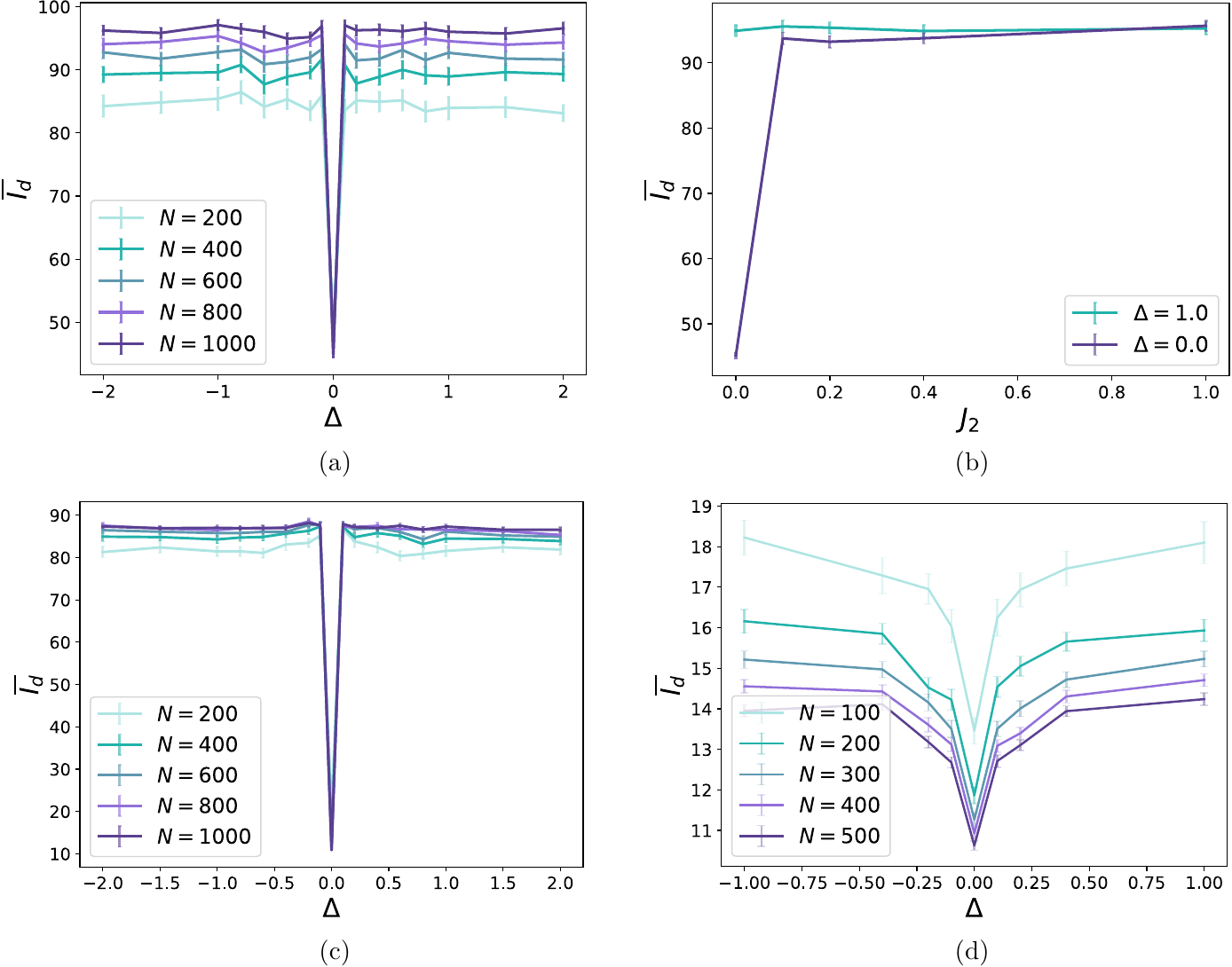}

\caption{Stationary intrinsic dimension $\overline{I_d}$ obtained by
averaging over $N$ QT at late times for the many-body models described in
    TABLE~\ref{tab:many-body}: (a) Model A with $J_1=1,\,\gamma_0=1$
    for $L=8$, as a function of $\Delta$, with $J_2=0$. For $\Delta=0$
    the Lindblad equation is Yang-Baxter integrable. (b) Model A with
    $J_1=1,\,\gamma_0=1$ for $L=8$, as a function of $J_2$, with fixed
    $\Delta=0.0,\,1.0$ and computed for $N=800$. For $\Delta=0=J_2$ the Lindblad equation is
    Yang-Baxter integrable. (c) Model B with $\gamma_1=1$. For
    $\Delta=0$ the model is Yang-Baxter integrable. (d) Model D with
    $\gamma_2=1$ from a state with fixed spin. For $\Delta=0$ the
    BBGKY hierarchy decouples.}
\label{fig: Idft XXZ}
\end{figure*}

\textcolor{black}{Finally, note that the Euclidean distance in Eq.~(13) is a convenient computational choice, but it is not necessarily unique. 
For data points sampled from a smooth finite-dimensional manifold, local intrinsic-dimension estimates are expected to be invariant under changes of Riemannian metric unless the data set is too sparse or noisy, since such distances are locally equivalent and preserve the volume scaling of small metric balls~\cite{Facco2019}. 
Another natural alternative that requires a further comment would be the Bures distance, because it acts on rays in the projective Hilbert space and not on wavefunction representatives. We have checked that our results are robust under this metric choice}~\footnote{\textcolor{black}{
For normalized pure states, the Bures distance is equivalent to the Euclidean distance minimized over the global $U(1)$ phase.
If the chosen wavefunction representatives define a smooth lift of the projective dataset, this phase can contribute at most one additional local direction to the Euclidean intrinsic dimension. 
For finite noisy datasets, however, a rapidly varying or effectively random global phase might also modify nearest-neighbor relations. 
We therefore performed a direct check for a representative example, namely the XXZ chain of Model~(A) in Fig.~\ref{fig: Idft XXZ}~(a). 
The results obtained with the Bures distance are compatible with the Euclidean-distance results within error bars.
}}


\begin{figure*}[t!]
\centering
\includegraphics[width=1\textwidth]{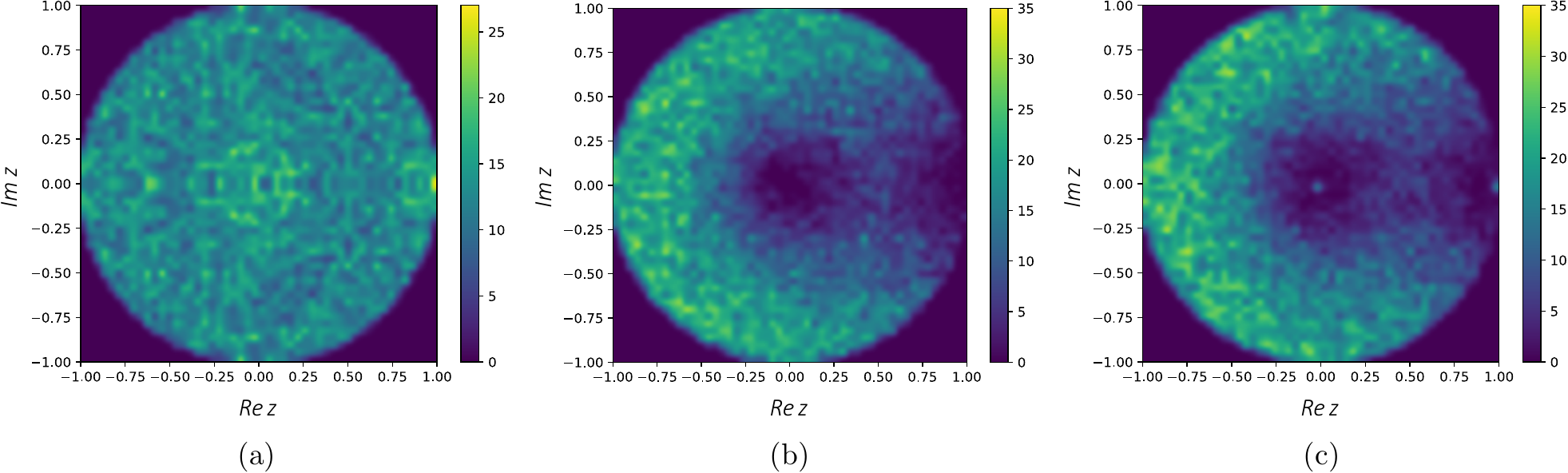}

 \caption{Complex ratios of the GHS kicked top {with parameters $S=80,\,\omega_z=4,\,g=5$}, in the regular and chaotic regimes: $(a)$ $\omega_x=0,\,k=0,\,\gamma=0.2$ (integrable); $(b)$ $\omega_x=0\,, k=4,\,\gamma=4$ (with kicks); $(c)$ for $\omega_x=2,\,k=0,\, \gamma=2$. The level repulsion is well captured by the angular distributions, having (a) $\overline{\cos\theta}=0.001$, (b) $\overline{\cos\theta}=-0.23$, (c) $\overline{\cos\theta}=-0.24$.} 

\label{fig: complex ratios}
\end{figure*}


\begin{figure}[t!]
\centering
\hspace*{-1cm}
\includegraphics[width=0.8\linewidth]{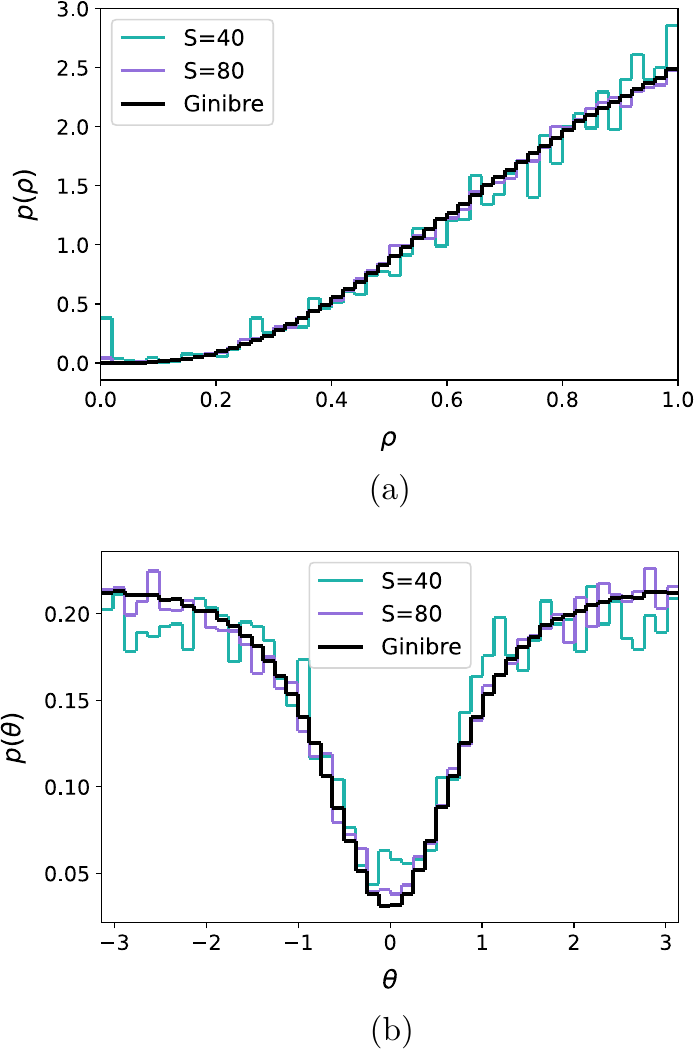}

 \caption{(a) Radial and (b) angular distributions corresponding to Fig.~\ref{fig: complex ratios} (c), showing the convergence of the complex ratios to chaotic spectral statistics. The reference distributions are obtained by sampling \textcolor{black}{$200$} $N\times N$ random matrices with size $N=10000$ from the Ginibre ensemble.} 

\label{fig: ginibre_convergence}
\end{figure}


\section{Driven-dissipative quantum top}
\label{sec: quantum top}

A prototypical model of quantum chaos is the kicked top, which has been extensively studied in the context of closed, unitary systems (e.g. \cite{Haake1987, Fox1994, Jacquod2001, Bandyopadhyay2004, Chaudhury2009, Neill2016, Bhosale2018, Sieberer2019, Vallini2024}). The effects of dissipation on spectral statistics were investigated in \cite{Grobe1988}. We begin our analysis revisiting the results of Ref.~\cite{Grobe1988} and extending them to a case where chaotic dynamics emerges purely from quantum fluctuations.
%
%
%
We start by recalling the form of the Hamiltonian, \emph{cf.} Eq.~\ref{eq: H quantum top}
\begin{equation*}
    \hat{H}(\omega_z,g,\omega_x,k) = \hat{H}_0 + \!\! \sum_{n=-\infty}^\infty \!\!  \delta(t-n\tau)\, \hat{H}_k\,,
\end{equation*}
with
$\hat{H}_0 = \omega_z \hat{S}_z + (g/S) \hat{S}_z^2 + \omega_x \hat{S}_x$ and 
$\hat{H}_k = (k/S) \hat{S}_y^2\,$.

The top is subject to a periodic driving, chosen for simplicity as impulsive kicks at times $t=n\tau$.
It is well-known that the top can be viewed as a model of many spins undergoing infinite-range collective processes by identifying the operators $\hat{S}_\alpha$ with the components of the total angular momentum operator of $N=2S$ individual spins 1/2. In this case the $1/S$ scaling with the total spin is the Kac factor of the associated long-range model \cite{Defenu2023}. Thanks to this representation the model can experimentally realized for instance with cold atoms \cite{Chaudhury2009} or superconducting qubits \cite{Neill2016}.
{\color{black} In the large-S limit,} the spectral statistics of the Hamiltonian in Eq.~\eqref{eq: H quantum top} exhibits a crossover from Poisson to Wigner-Dyson as the kick strength $k$ is increased, which is a signature of unitary quantum chaos \cite{Haake1987}. 
Another important property is that the quantum top becomes semiclassical when the spin $S$ is large, \emph{cf.} Appendix~\ref{app: semiclassical limit}. In this limit, $1/S$ acts as an effective Planck's constant that bounds quantum fluctuations. In agreement with the BGS conjecture, the presence of chaos signaled by the spectral statistics mirrors the emergence of chaos in the limiting classical model, whose phase space displays growing chaotic regions in presence of driving \cite{Bohigas1984, Haake2010}.


\begin{figure*}[t!]
\centering

\includegraphics[width=0.95\textwidth]{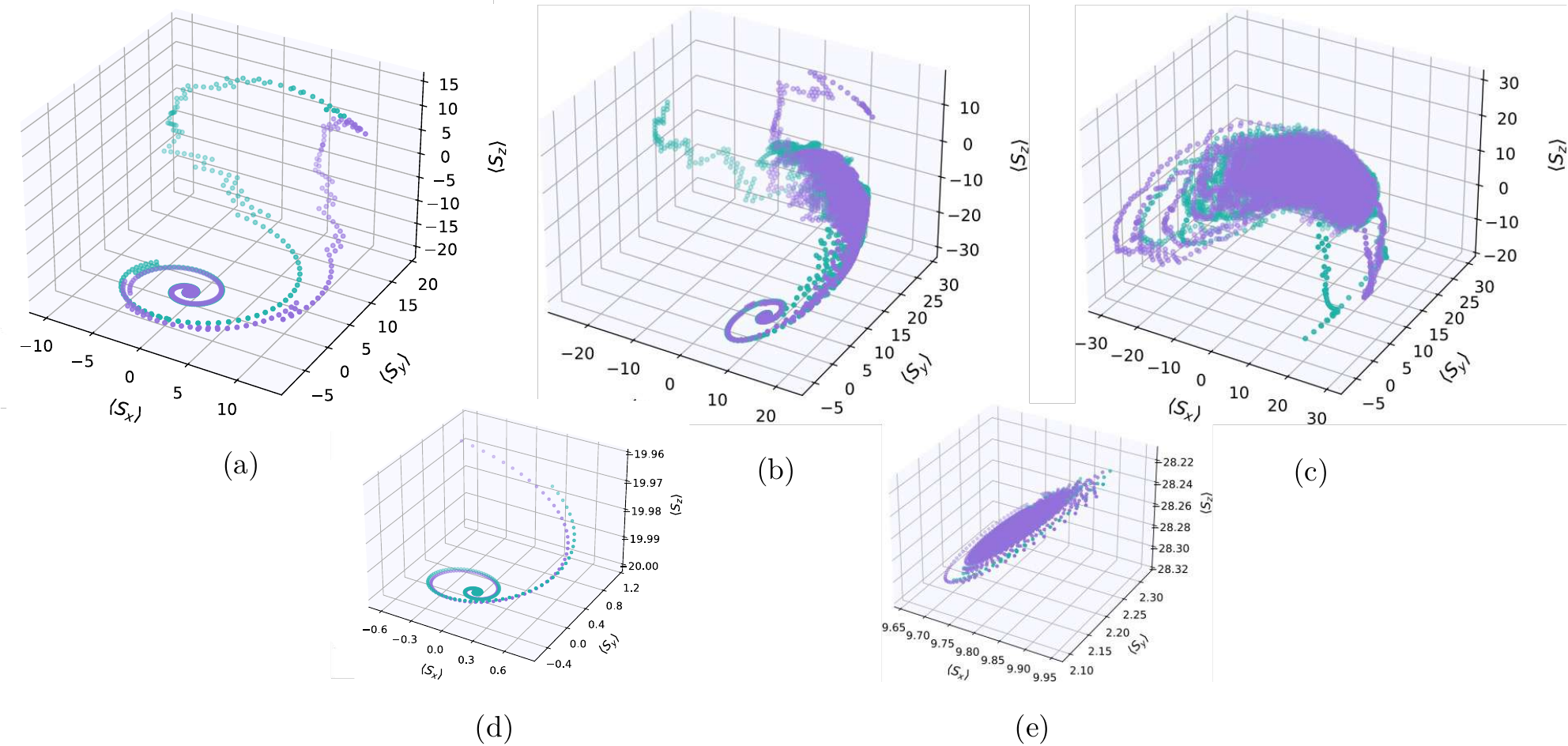}
\caption{Chaotic behavior of the dissipative quantum top as seen from individual QT. We plot two QT of the model given by Eq.~\eqref{eq: lindblad quantum top} with $\omega_z=1, \,g=5,\,k=0$, considering in (a), (b) $\omega_x=0$, in (c), (d) $\omega_x=3$ and in (e) $\omega_x=5$. Even if trajectories are always perturbed by stochastic effects, the local structure between consecutive quantum jumps is one-dimensional in the integrable case (a). The subfigure (d) is a zoom on the spiraling region at the end of the trajectories in (a), showing that different trajectories are attracted towards the same one-dimensional region. Increasing $\omega_x$, a transient chaotic region starts to appear, where the trajectories get trapped before escaping to another long-time attractor. As shown in (e), this structure is not one-dimensional, in contrast to the integrable case. 
For sufficiently large values of $\omega_x$, the chaotic attractor takes over, as we depict in (e). The appearance of initially small chaotic regions that become dominant increasing the integrability-breaking coupling is reminiscent of what happens in the classical systems, where chaotic trajectories take over the phase space.
} 
\label{fig: trajectories}
\end{figure*}

We now introduce dissipation through a jump operator $\hat{L}=\sqrt{\gamma}\hat{S}_-$. This gives rise to
the Lindblad equation
\begin{equation}
\label{eq: lindblad quantum top}
    \dv{\hat{\rho}}{t} = -i[\hat{H},\hat{\rho}] + \frac{\gamma}{S}\Bigl(\hat{S}_-\hat{\rho} \hat{S}_+ - \frac12\{\hat{S}_+\hat{S}_-,\hat{\rho}\}\Bigr)\,.
\end{equation}
Both the Hamiltonian and the dissipator commute with the total angular momentum $\hat{S}^2 = \hat{S}_x^2+\hat{S}_y^2+\hat{S}_z^2$, which therefore defines a {strong symmetry}, and the dynamics takes place in a subspace with conserved spin. 
%
%
The presence of dissipative chaos can be detected by analyzing the spectrum of the Lindblad generator, as discussed in Section~\ref{ssec: linbdlad integrability}. For a Floquet model like Eq.~\eqref{eq: H quantum top}, one studies the eigenvalues of the stroboscopic time-evolution operator. Vectorizing the Eq. \eqref{eq: lindblad quantum top} leads to 
\begin{equation}
    \check{\mathcal{L}} \equiv \check{\mathcal{L}}_0 + \!\! \sum_{n=-\infty}^\infty \!\! \delta(t-n\tau)\,\check{\mathcal{L}}_{k} \,, 
\end{equation}
where $\check{\mathcal{L}}_0 = -i\check{\mathcal{H}}_0 + \check{\mathcal{D}}$ includes the time-independent part of the Hamiltonian and the dissipator, while $\check{\mathcal{L}}_k = -i\check{\mathcal{H}}_k$ only the kicks. 
Defining $|\rho(t+\tau)\rrangle = \hat{\mathcal{F}}_\tau|\rho(t)\rrangle$, the evolution operator for a single period is
\begin{equation}
    \hat{\mathcal{F}}_\tau = e^{\tau\hat{\mathcal{L}}_0}e^{\hat{\mathcal{L}}_k} = e^{\tau(\hat{\mathcal{D}}-i\hat{\mathcal{H}}_0)}e^{-i\hat{\mathcal{H}}_k}\,.
\end{equation}
We focus on two scenarios: 
\begin{itemize}
\item{} We fix $\omega_x=0$ and vary the kick strength $k$.
\item{} We set $k=0$ and tune $\omega_x$. 
\end{itemize}


 \begin{figure}[t!]
\centering

    \includegraphics[width=0.95\columnwidth]{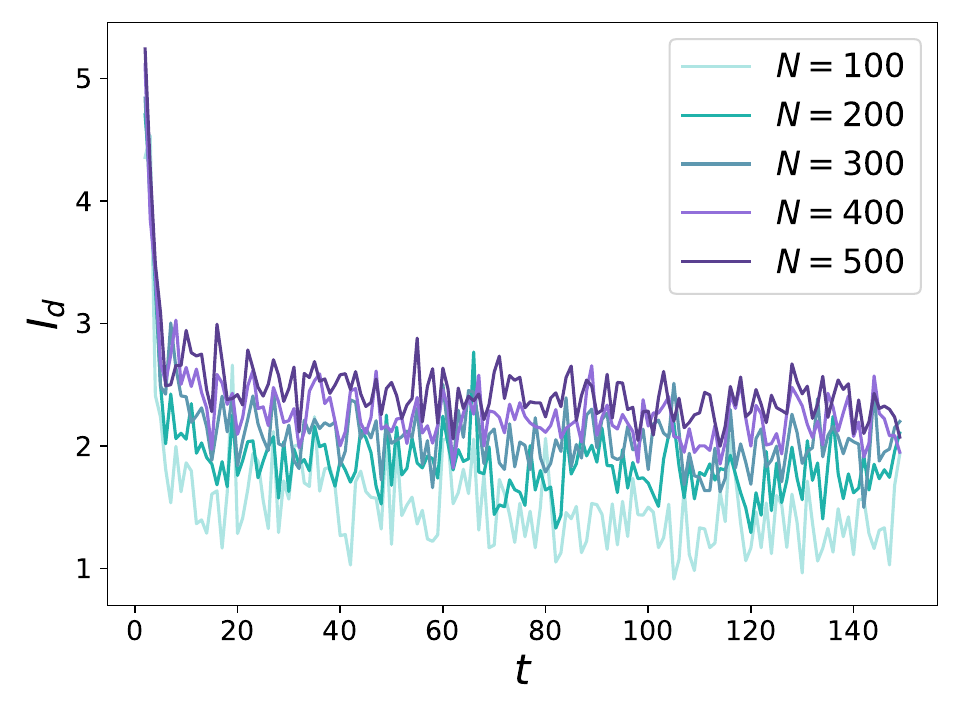}
 
\caption{Time evolution of the intrinsic dimension for the dissipative top with $\omega_z=1,\,g=5,\,\omega_x=4.0,\,k=0$ and $\gamma=2$, $S=30$. The value of $I_d(t)$ is estimated using different different numbers of trajectories, encoding now the scale dependence of $I_d$. With a larger $N$, the points are on average closer to each other and probe a finer scale of the data set, resulting in increasing values like for the $I_d$ calculated along individual trajectories.} 
\label{fig: Idft dynamics}
\end{figure}


\begin{figure*}[t!]
\centering

    \includegraphics[width=0.9\textwidth]{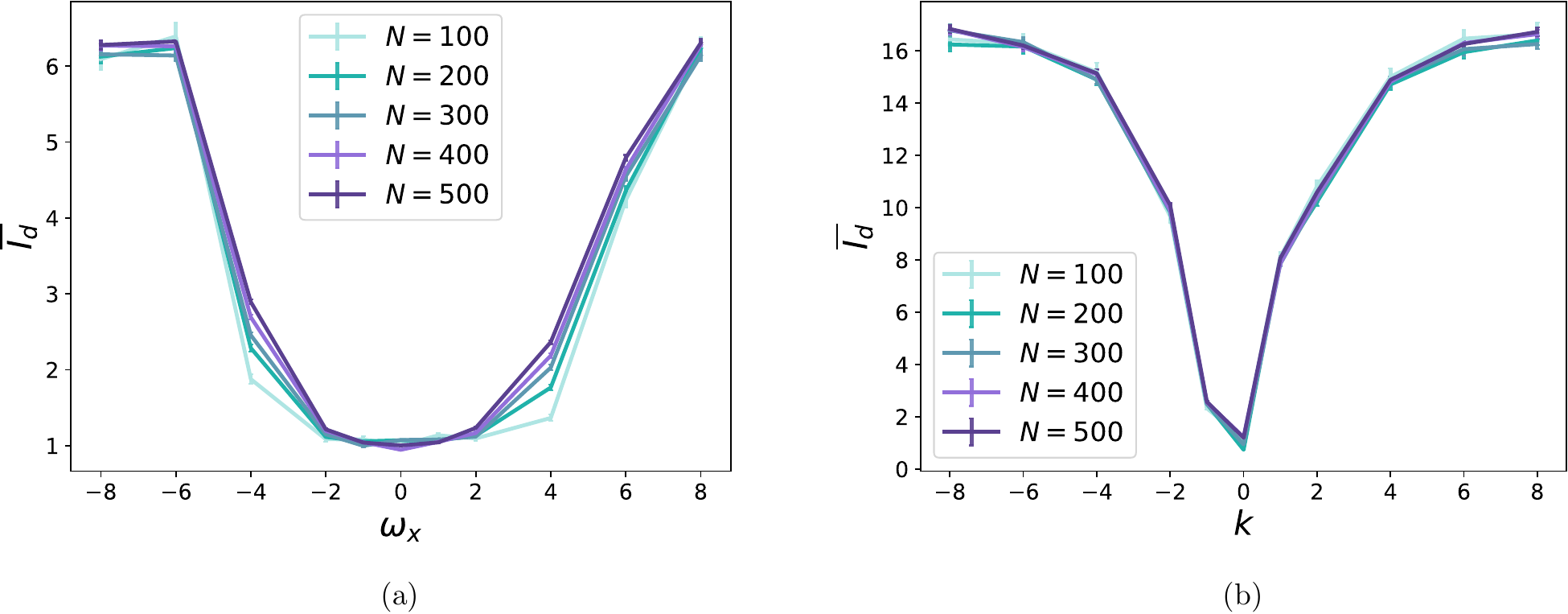}

\caption{Late-time average of $I_d(t)$ for the dissipative top with $\omega_z=1,\,g=5,\,\gamma=2$ and $S=30$, varying respectively $\omega_x$ and $k$ in (a), (b).  The error bars shown represent the standard deviation of $I_d(t)$ for late times. The integrable point of the quantum top is associated to one-dimensional trajectories; when chaos is present the $I_d$ increases in a model dependent manner.
}
\label{fig: Idft top}
\end{figure*}

The corresponding distributions of the complex ratios defined in Eq.~\eqref{eq: complex ratio} are reported in Fig.~\ref{fig: complex ratios}. 
When $k=0$ and $\omega_x=0$ the model is integrable \cite{Ribeiro2019} and correspondingly shows signatures of 2d-Poisson statistics, \emph{cf.} Fig.~\ref{fig: complex ratios} (a).
Adding kicks leads to chaotic behavior and complex ratio distributions described by the Ginibre ensemble, \emph{cf.} Fig.~\ref{fig: complex ratios} (b) and (c).
This was first studied in Ref.~\cite{Grobe1987}, where the Poisson-to-Ginibre crossover, {\color{black} on increasing $S$}, was associated with the emergence of chaos as the kick strength $k$ is increased.

Including $\omega_x$ makes the model quantum chaotic without introducing chaos in its classical limit~\footnote{\textcolor{black}{In the quantum case, we expect a finite $\omega_x$ to break the integrability of the model, since the term proportional to $\hat S_x$ in the Hamiltonian breaks the weak symmetry $\mathcal{L}([\hat S_z,\hat\rho]) = [\hat S_z,\mathcal{L}\hat\rho]$, on which the proof of Ref.~\cite{Ribeiro2019} relies. The analysis of the spectrum confirms this expectation.}}. As we describe in the Appendix~\ref{app: semiclassical limit}, the classical limit of Eq.~\eqref{eq: lindblad quantum top} is a dynamical system on the 2-sphere, and, in absence of kicks, chaos is topologically forbidden by the Poincaré-Bendixson theorem. Only the inclusion of a driving term can generate chaotic orbits \cite{Wiggins2003, Strogatz2015}. 
The corresponding distribution of the eigenvalues shows signatures of chaos, even when the behavior of the classical limit remains regular, therefore providing an exception to the GHS conjecture. We include in Fig.~\ref{fig: ginibre_convergence} the radial and angular distributions to further support our claims.
This result also implies that a direct quantum analog of the Poincaré-Bendixson theorem does not exist. 
Our observations are corroborated by recent literature \cite{Villasenor2024, Ferrari2025}, which support the idea that in open systems quantum chaos can arise even in the absence of its classical counterpart.

Before calculating the intrinsic dimension, it is instructive to consider the trajectories on the 3d space spanned by the expectation values $\langle \hat{S}_\alpha \rangle\,,\,\,\alpha=x,y,z$  evaluated in a single QT, as shown in Fig.~\ref{fig: trajectories}. For parameter values corresponding to integrable Lindblad equations the underlying one dimensional structure of these trajectories is evident.  For values of $t$ between consecutive quantum jumps the discrete points $\langle S_\alpha\rangle_t$ are aligned along a curve, even though there are random discontinuities introduced by the jumps. When quantum chaos is present according to the complex level spacing analysis, the trajectories appear to “fold upon themselves” and the time series of $\langle S_\alpha \rangle_t$ fills a region of higher dimensionality of the space. Remarkably, this feature is present when chaos is introduced by kicks as well as and when $H$ is kept time-independent, in agreement with the spectral statistics but not with the classical limit.
Counterexamples to the GHS conjecture have been employed {in the literature} to question the validity of the \mbox{level-spacing statistics 
criterion for dissipative} chaos~\cite{Villasenor2024}. {Our results instead support the picture arrived at by means of the level-spacing statistics analysis},
indicating that it is the quantum–classical correspondence that fails in this regime.
This happens because the regular classical dynamics can describe the model at finite $S$ only up to some time scale analogous to the Ehrenfest time. Beyond that, the $O(1/S)$ fluctuations bring the state in a fully quantum regime, where the complete spectrum has a significant influence on the dynamics. 

The expectation values $\langle \hat{S}_\alpha \rangle_t$  only provide a small part of the information encoded in the QT.
To understand whether the structure of chaotic trajectories is intrinsically higher-dimensional  we have to consider the data set formed by the full QT $|\psi(t)\rangle \in \mathcal{H}$.
As wave functions in a high-dimensional Hilbert space cannot be easily visualized we focus on calculating the intrinsic dimension. As discussed in Sec.~\ref{ssec: data sets}, we calculate $I_d$ as a function of time using the data sets $\mathcal{D}_t \equiv \{|\psi_t\rangle_{\mathcal{N}_1},\,|\psi_t\rangle_{\mathcal{N}_2},\dots,\,|\psi_t\rangle_{\mathcal{N}_N}\}$, consisting of the states of each of the $N$ trajectories at the same time $t$. 
The results are shown in Fig.~\ref{fig: Idft top}. As we have seen in Fig.~\ref{fig: trajectories}, at early times trajectories spread over the Hilbert space along independent directions. This corresponds to the initial large-$I_d(t)$ regime in Fig.~\ref{fig: Idft dynamics}. At later times, the trajectories get trapped in a complex subset of $\mathcal{H}$, that could be interpreted as an attractor. At any sufficiently late time $t$, the points $\{|\psi_t\rangle_{\mathcal{N}_i}\}$ are randomly drawn from this subset, implying that we can obtain an estimate of the dimensionality of the attractor as a late time average 
\begin{align}
\overline{I_d}=\frac{1}{|T-\bar{t}|}\int_{\bar{t}}^T \,I_d[\mathcal{D_t}]\,dt\,,
\end{align}
where $I_d[\mathcal{D}_t]$ is estimated from Eq.~\eqref{eq: Id fit} and $\bar{t}$ is the time at which its values become stationary. 
In Fig.~\ref{fig: Idft top} we show the results for $\overline{I_d}$ as functions of $\omega_x$ and $k$. The results corroborate the intuitive picture suggested by the time evolution of the expectation values $\langle \hat{S}^\alpha\rangle$: close to the integrable point we estimate $\overline{I_d}\approx1$ and the intrinsic dimension increases as we increase integrability-breaking couplings, meaning that individual QT describe a data manifold with increasing complexity. Including in the model further dissipation channels induced by $\hat{S}_+$ or $\hat{S}_z$ would not break integrability \cite{Ribeiro2019}, and the associated QT remain one-dimensional. Also the fluctuations, measured by the standard deviation of the late-time values of $I_d$, are  very small when the model is integrable.

\begin{figure}[t!]
\centering

\includegraphics[width=0.4\textwidth]{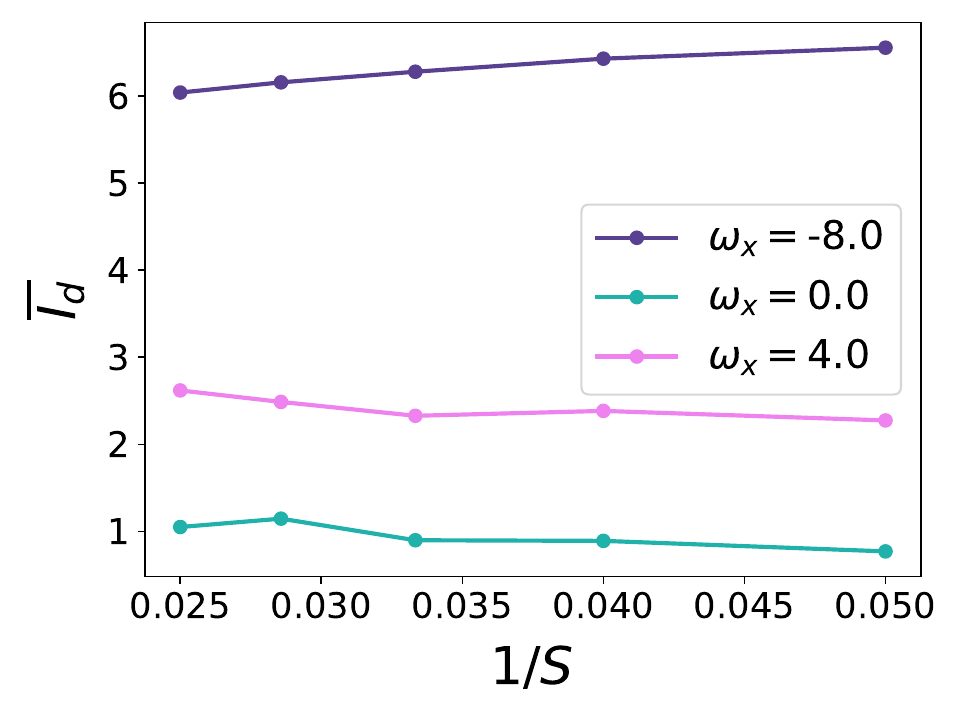}

\caption{\textcolor{black}{Dependence of the $I_d$ of the autonomous quantum top on the Hilbert-space dimension through the total spin size. Different colors correspond to different values of the integrability-breaking coupling.}}
\label{fig: Idft scaling S}
\end{figure}
%

{\color{black}
Despite the fact that at the integrable point the intrinsic dimension is minimal, its growth is rather smooth on increasing the integrability-breaking term. This should be contrasted with the many-body case discussed in the next section, where the separation between integrable and chaotic regimes is sharper. We expect that in this case, due to the linear increase of Hilbert space dimension with
$S$, finite-size effects are more severe and one should go to larger $S$ to see a sharpening of the curves around
the integrable point.
This motivated the finite-size analysis shown in Fig.~\ref{fig: Idft scaling S}. 
Increasing \(S\) enlarges the embedding space, whose real dimension is \(D=4S+2\). 
Within the range of values considered, however, the dependence of \(I_d\) on \(S\) remains weak on the scale of the plotted values. 
At the integrable point \(\omega_x=0,\,k=0\), one obtains \(I_d\simeq 1\) independently of the value of the spin, while at finite integrability-breaking coupling the curves show only mild increases or decreases depending on the parameter choice. 
The available data therefore do not allow us to extract a reliable asymptotic extrapolation.

Finally, in the autonomous case the immediate vicinity of \(\omega_x=0\) requires an additional comment, because at finite \(S\), we observe a shallow interval in which \(I_d\) remains close to its minimum value. As discussed in Section~\ref{sec: quantum top}, we expect only \(\omega_x=0\) to be integrable, but we have no statistical evidence of quantum signature of chaos in the spectra for small values of \(\omega_x\), as spectral distributions are known to exhibit finite size crossovers in the vicinity of integrable points~\cite{Rabson_2004}. A similar crossover, possibly enhanced by the regular semiclassical limit, may be affecting the intrinsic dimension as well. Consequently, the finite-size data do not allow us to rule out residual integrable effects for sufficiently small nonzero \(\omega_x\), even though the minimum of \(I_d\) occurs at the expected point.
}

The values of $I_d$ show a tiny variation with the number of trajectories, especially at non-integrable points. This is an effect of the scale dependence of the intrinsic dimension, that we study in detail in Appendix~\ref{app: Id scaling}. 
Calculating the $I_d$ along single trajectories gives results consistent with the above analysis, as discussed Appendix~\ref{app: Id single traj}.


%

%
\begin{figure*}[]
\centering

    \includegraphics[width=\textwidth]{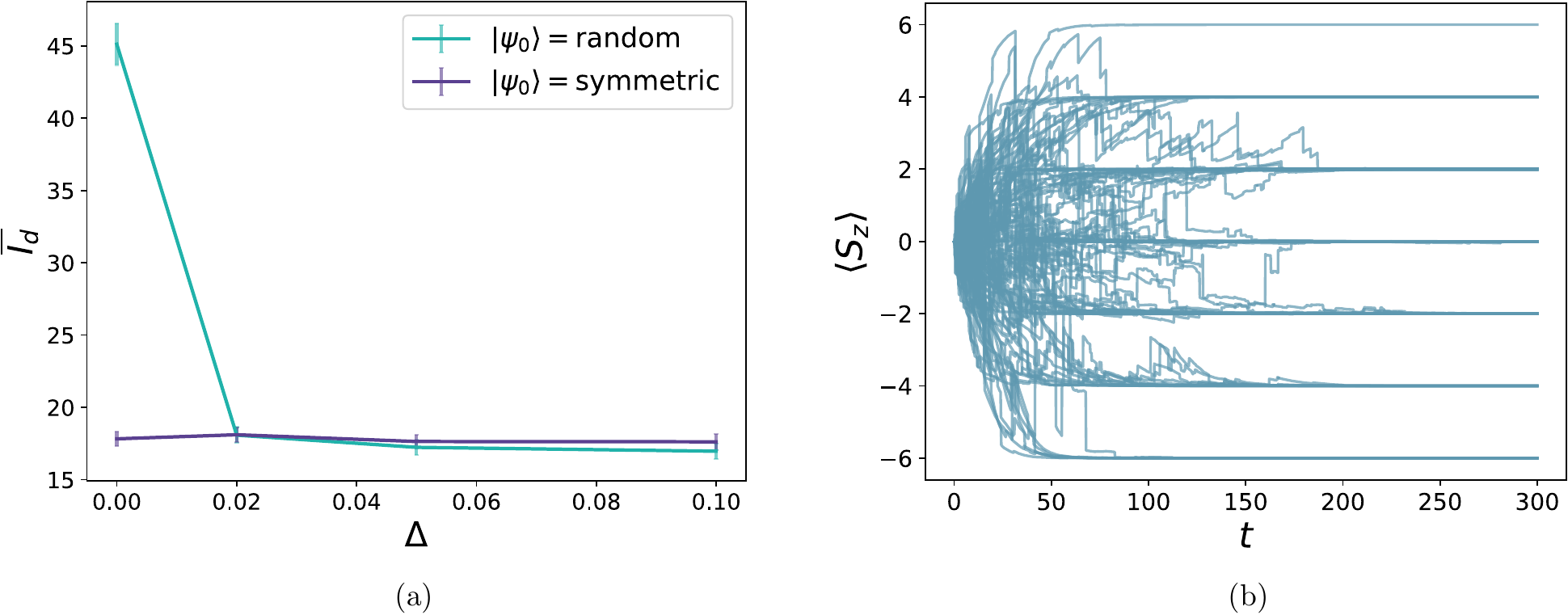}

\caption{Dissipative freezing mechanics as seen by: (a) the intrinsic dimension; (b) the expectation values of the charge. 
}
\label{fig: diss freezing}
\end{figure*}

%

\begin{figure}[t]
\centering

\includegraphics[width=0.45\textwidth]{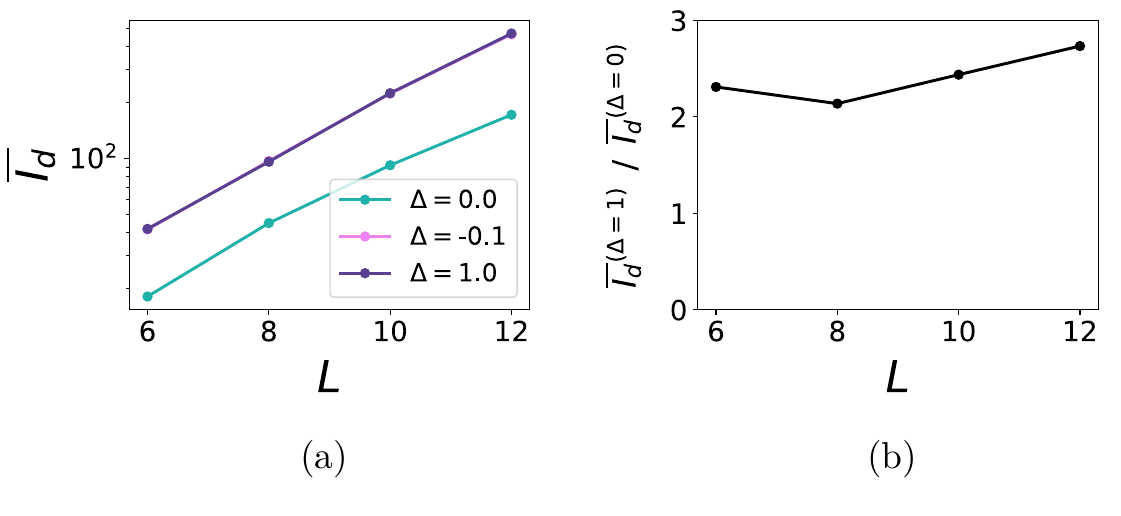}

\caption{\textcolor{black}{Dependence of the $I_d$ of the XXZ chain with dephasing (Model (A)) on the Hilbert-space dimension through the chain length $L$ (note the logarithmic scale). (a) Different colors correspond to different anisotropies $\Delta$; non-integrable values are superimposed. (b) Ratio between the integrable and non-integrable values, signaling a slight deepening of the minimum as the sizes increase.}}
\label{fig: Idft scaling L}
\end{figure}
%

\section{Quantum spin chains with dissipation}
\label{sec: many body}

We now turn to the case of genuine many-body models using
variations of the spin-1/2 Heisenberg XXZ chain with dissipation as
our testing ground. This will allow us to study the effects of quantum
integrability, the decoupling of the open system analog of the BBGKY
hierarchy for correlation functions and strong symmetries on the
intrinsic dimension of QT.
\subsection{Models}
\label{ssec: models many body}

We start by recalling the models introduced in Table~\ref{tab:many-body}. 
The many-body models we consider in the following are summarized in
TABLE~\ref{tab:many-body}. Their unitary terms are particular cases of
the spin-1/2 Hamiltonian in Eq.~\eqref{Hamiltonian} 
\begin{align}
\hat{H}(J_1,J_2,\Delta) &= \sum_{j=1}^{L-1}J_1 \Bigl[\hat{\sigma}^x_j\hat{\sigma}^x_{j+1} + \hat{\sigma}^y_j\hat{\sigma}^y_{j+1}\Bigr] + \Delta\,\hat{\sigma}^z_j\hat{\sigma}^z_{j+1}\nonumber\\
    &\,+J_2\sum_{j=1}^{L-2}\Bigl( \hat{\sigma}^x_j\hat{\sigma}^x_{j+1} + \hat{\sigma}^y_j\hat{\sigma}^y_{j+1}\Bigr)\ .
\label{Hamiltonian_recall}
\end{align}
%
For $J_2=0$ this reduces to the integrable spin-1/2 Heisenberg XXZ chain. The Lindblad equations are chosen by combining (\ref{Hamiltonian_recall}) with specific choices of jump operators:
\begin{itemize}
\item{} Model A: Local dephasing
\begin{equation}
\label{eq: dephasing}
\hat{L}^{(0)}_j=\sqrt{\gamma_0}\,\hat{\sigma}^z_j\,.
\end{equation}
The resulting Lindblad equation is non-integrable as can be verified
by a spectral statistics analysis~\cite{Akemann2019,Sà2020}, except at
the special point $\Delta=0$ and $J_2=0$. Yang-Baxter integrability at
$(\Delta,J_2)=(0,0)$ has been proven by constructing the vectorized
representation of the Lindbladian, which becomes a non-Hermitian,
integrable Fermi-Hubbard model with imaginary interaction proportional
to $i\gamma_0$ \cite{medvedyeva2016exact}.
At the integrable point $(\Delta,J_2)=(0,0)$ the model
features a decoupled BBGKY hierarchy as well. Tuning $J_2$ away from
zero retains this feature, while breaking integrability.
\item{} Model B: $J_1=J_2=0$ and jump operators
\begin{equation}
\label{eq: frag}
    \hat{L}^{(1)}_j = \frac{\sqrt{\gamma_1}}2\left(\boldsymbol{\hat{\sigma}}_j\cdot\boldsymbol{\hat{\sigma}}_{j+1}+1\right)\,,
\end{equation}
where $\boldsymbol{\hat{\sigma}}=(\hat{\sigma}_x,\hat{\sigma}_y,\hat{\sigma}_z)$. The purely
dissipative model $\Delta=0$ is known to be Yang-Baxter integrable
\cite{ziolkowska2020yang}, while $\Delta\neq 0$ breaks
integrability. 
\item{} Model C: $J_2=\Delta=0$ and
\begin{eqnarray}
\label{eq: gamma_h}
\hat{L}_j^{(0)} &=&\sqrt{\gamma_0}\hat{\sigma}^z_j\,\nonumber\\
\hat{L}_j^{(2)} &=& \sqrt{\gamma_2}\,\hat{\sigma}^+_j\hat{\sigma}^-_{j+1}\,  \\
\hat{L}^{(3)}_j &=& \sqrt{\gamma_3}\,\hat{\sigma}^-_j\hat{\sigma}^+_{j+1}\,, \nonumber
\end{eqnarray}
 This is relevant because, as we will see, changing the dissipator can have subtle effects related to a strong symmetry.
\item{} Model D: $J_2=\Delta=0$ and
\begin{equation}
\label{eq: gamma_h}
 L_j^{(2)} = \sqrt{\gamma_2}\,\hat{\sigma}^+_j\hat{\sigma}^-_{j+1}=(L^{(3)}_j)^\dagger.
\end{equation}
This corresponds to a generalization of the Quantum Symmetric Simple Exclusion Process \cite{Bernard2019Open,essler2020integrability,barraquand2025introduction}, which is known to exhibit a decoupled BBGKY hierarchy \cite{Penc2025}.
\end{itemize}

\subsection{Results}
\label{ssec: results many body}

The picture that emerged in Section~\ref{sec: quantum top} is that the 
integrable points of the dissipative quantum top are characterized by
an intrinsic dimension $I_d\approx1$, which increases as chaotic
behavior sets in. We now examine how these findings extend to more
general settings, separating the discussion of the effects of
integrability from those arising from the decoupling of correlation
functions and dissipative freezing. 

\begin{itemize}

\item[$\circ$]{ \emph{Signatures of integrability}}

\end{itemize}

We start from the XXZ chain subject to dephasing (\textcolor{black}{Model (A)}).
QT are generated starting from a fixed random state, and we calculate
the intrinsic dimensions as a function of time.  
The initial transient is generally faster and more difficult to access
than in the single-body case, and  $I_d(t)$ fluctuates almost
immediately around its late-time value.  
As shown in Fig.~\ref{fig: Idft XXZ}, the late-time average of the
intrinsic dimension clearly indicates the presence of a quantum
integrable point as the point of minimal complexity. We have
investigated the evolution of $\overline{I_d}$ to switching on either
an exchange anisotropy $\Delta$ or a next-nearest-neighbor hopping
$J_2$, which respectively break the integrability of only the
Lindbladian, and of both the Lindbladian and its Hamiltonian part. Both
perturbations quickly drive $\overline{I_d}$ to a constant larger
value characteristic of the non-integrable regime. Note that this
value is unaltered when $\Delta$ and $J_2$ are both switched on,
indicating that $\overline{I_d}$ does not simply react to any
insertion of a new operator generating the dynamics.  
%
%

This example of the XXZ chain highlights the importance of calculating
the intrinsic dimension on the Hilbert space rather than focusing on a
small number of expectation values along QT, \emph{cf.} Fig.~\ref{fig:
  trajectories}. The results shown in Fig.~\ref{fig: 
  Idft XXZ} are obtained for a short chain with $L=8$
sites. The intrinsic dimension at the integrable point
is $\overline{I_d}\approx50$, which is much smaller than the embedding dimension
$D=2^{9}$, 
but still too large to be captured visually in a simple
way. To fully exhibit the drop in complexity of QT by
analyzing expectation values would require the identification of
approximately 50 appropriately chosen operators, which is a highly
nontrivial task in models without a simple classical limit.

%
In this case, increasing the length of the chain
results in a larger embedding space with dimension $D=2^{L+1}$, and, as reported in Fig.~\ref{fig: Idft scaling L}, the intrinsic dimension correspondingly
increases exponentially for all values of $\Delta$, showing no
significant differences in the scaling law between integrable and
non-integrable points.

The effects shown in Figs.~\ref{fig: Idft XXZ} (a) and
(b) could have their origin in integrability or BBGKY
decoupling, since the correlation functions also decouple at
$\Delta=0\,,\,\,J_2=0$.
To separate the effects of integrability and BBGKY decoupling we
consider Model (B), which exhibits an integrable point with coupled
hierarchy of correlation functions. In Fig.~\ref{fig: Idft XXZ} (c) we
show the results for $\bar{I_d}$ as a function of the anisotropy
parameter $\Delta$. We see that $\bar{I_d}$ drops sharply at the
integrable point $\Delta=0$.
In contrast to the quantum top, at these integrable points
$\overline{I_d}$ is significantly larger than $1$. This
is expected, because in many-particle systems quantum integrability
imposes only an extensive number of local constraints on the dynamics.


\begin{itemize}

\item[$\circ$]{\emph{Signatures of decoupling and dissipative freezing}}

\end{itemize}

\begin{figure*}[t!]
\centering

    \includegraphics[width=\textwidth]{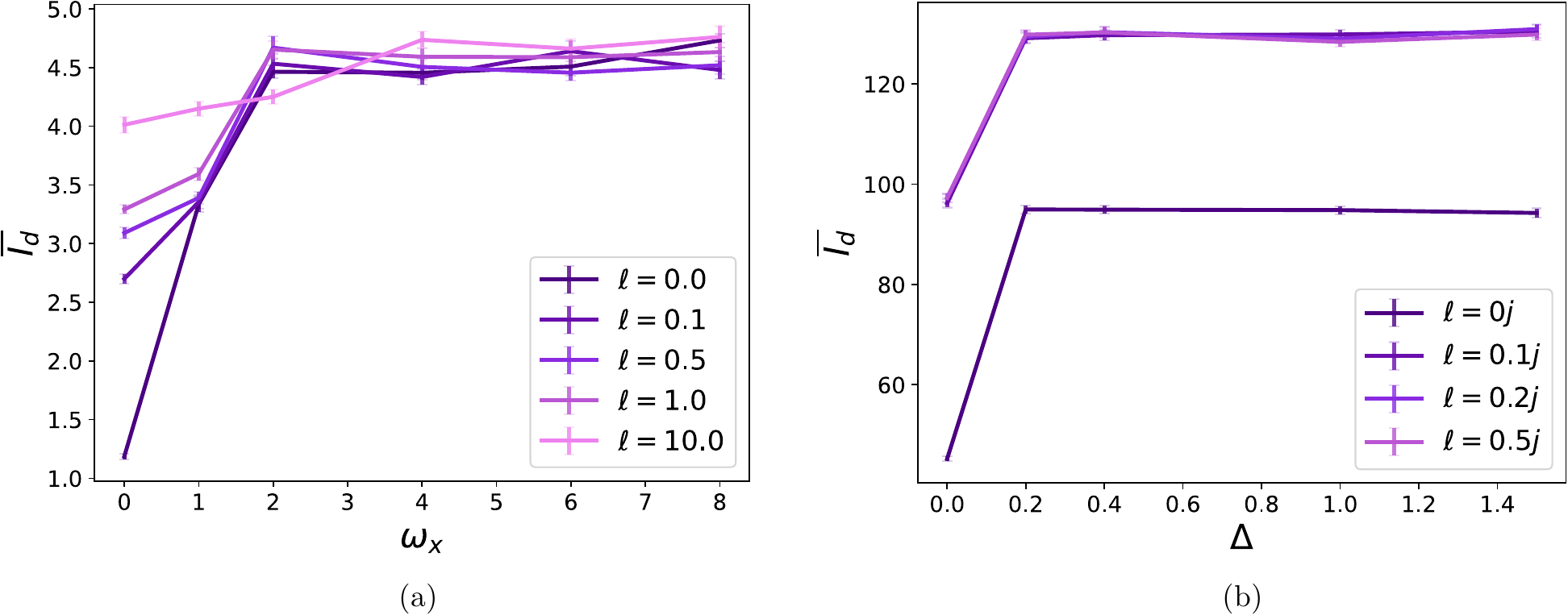}

\caption{Intrinsic dimension of quantum trajectories generated by different unravelings. The $I_d$ is always calculated at fixed times. In (a) we consider the quantum top of Eq.~\eqref{eq: lindblad quantum top} where integrability is broken by $\omega_x>0$, and for simplicity we have taken $\ell$ real. The results for the XXZ chain with dephasing defined are presented in (b), where we have chosen instead an imaginary $\ell$, because it resulted in a cleaner Pareto distributions.} 

\label{fig: Idft modQJ}
\end{figure*}


What we have observed so far does not exclude the possibility that the
BBGKY decoupling by itself could affect the complexity of QT. Before
specifically addressing this question, we gradually turn on the dissipator in
Eq.~\eqref{eq: gamma_h} from the integrable XX chain with
$\Delta=0\,,\,\,J_2=0$ already subject to dephasing
(\textcolor{black}{Model (C)}). As reported in Fig.~\ref{fig: diss 
  freezing}, starting from a random state, we observe that adding a
finite $\gamma_2>0$ sharply decreases the intrinsic dimension. At a
first glance this result seems challenging to interpret, as $\gamma_2$
is an integrability-breaking coupling. This is caused by another
effect that reduces the complexity of QT, which had been called
\emph{dissipative freezing}~\cite{Munoz_2019, Tindall_2022}. This
model has a strong $U(1)$ symmetry generated by the total
magnetization $\hat{S}_z=\frac{1}{2}\sum_{j=1}^{L}\hat{\sigma}^z_j$, commuting
separately with Hamiltonian and jump operators. Strong symmetries are associated with conservation laws on average and preserving
$\text{tr}[\hat{\rho}\hat{Q}]$ for some charge $Q$ is not sufficient to enforce the
conservation law along individual trajectories. After a quantum jump
induced by an operator $\hat{L}$, one has 
\begin{equation*}
    \langle \hat{Q}\rangle_t\,\to\,\langle \hat{Q}\rangle_{t+dt}=\frac{\bra{\psi_t}\hat{L}^\dagger \hat{Q}\hat{L}\ket{\psi_t}}{\bra{\psi_t}\hat{L}^\dagger \hat{L}\ket{\psi_t}}
\end{equation*}
and $[\hat{L},\hat{Q}]=0$ is not sufficient in general to preserve the expectation
values. Nevertheless, if $|\psi_t\rangle$ is an eigenstate of $Q$ with
eigenvalue $q$ one obviously obtains $\langle Q\rangle_t=q=\langle
Q\rangle_{t+dt}$, and analogously for the deterministic term. If a
stochastic trajectory happens to hit an eigenstate, the dynamics gets
trapped in the eigenspace and expectation values become frozen. While
this may seem unlikely, it is actually expected to happen under
generic conditions~\cite{Munoz_2019, Tindall_2022}. Instead, the
strong symmetry condition together with $(\hat{\sigma}^z)^2=\mathbb{I}$
imply that dephasing quantum jumps are unable to modify the
expectation values even if the state belongs to a superposition of
eigenspaces. Adding $\gamma_2>0$ triggers the dissipative freezing,
which becomes the main phenomenon witnessed by the intrinsic
dimension, therefore hiding the effect of the BBGKY decoupling. To remove
this spurious albeit interesting effect, we consider an initial state
with fixed spin.   
We note that in the previous models we have considered this would only
rescale all dimensionalities, since the dynamics becomes constrained
in an eigenspace. 

This brings us \textcolor{black}{Model (D)}, with nearest-neighbor
terms in the Hamiltonian and dissipation induced only by
Eq.~\eqref{eq: gamma_h}. As shown in Fig.~\ref{fig: Idft XXZ},
$\overline{I_d}$ detects again the point with structurally simplified
for dynamics with a local minimum at $\Delta=0$, where the model
exhibits a decoupled BBGKY hierarchy. We conclude that the intrinsic
dimension of QT is able to witness the complexity of Lindblad
dynamics, regardless of the mechanism driving the potential
simplification.

\section{Dependence on the unraveling}
\label{sec unraveling}

So far we have focused on the intrinsic dimension of quantum jump trajectories. Any Lindblad equation has infinitely many unravelings, it is therefore important to check whether our findings are a special property of quantum jumps, or if they are robust under modifications of the stochastic process. Consider the transformation
\begin{subequations}
\label{eq: lindblad gauge}
\begin{eqnarray}
    \hat{L}_\alpha\,&\to& \,\hat{L}_\alpha+\ell_\alpha\,,\\
    \hat{H}\,&\to & \,\hat{H}-\frac{i}{2}\sum_\alpha \gamma_\alpha \left(\ell_\alpha^*\hat{L}_\alpha - \ell_\alpha \hat{L}_\alpha^\dagger\right)+r\,,
\end{eqnarray}
\end{subequations}
where $\ell_a\in\mathbb{C}$ and $r\in\mathbb{R}$. This is a symmetry
of the Lindblad equation Eq.~\eqref{eq: lindblad}, and the evolution
of the mean state $\rho$ is unaffected by the transformation. However,
the stochastic processes defined by the quantum jump protocol over
Eq.~\eqref{eq: lindblad gauge} are different, providing infinite
unravelings of the same Lindblad equation. Other unravelings exist, 
such as homodyne or heterodyne QT (see Refs.~\cite{Wiseman2009,
  Jacobs2014}). We focus on Eq.~\eqref{eq: lindblad gauge} because the
parameters $\ell_a,\,r$ allow us to continuously deform the
unraveling.

In Fig.~\ref{fig: Idft modQJ} we show the intrinsic dimension for different
unravelings for the quantum top and the XXZ chain with dephasing. For
simplicity we have kept $r=0$ and for the spin chain the unraveling
has been modified considering $\ell_j=\ell$ equal at all sites. We see
that the values of $\overline{I_d}$ are in general unraveling-dependent, but for
all cases considered the integrable point has always remained the
point with minimal intrinsic dimension.
Varying $\ell$ for the quantum top results in approximately constant
intrinsic dimensions as long as the dynamics is chaotic. At the integrable
point $\overline{I_d}$ increases with $\ell$, while the overall range of
variation of $\overline{I_d}$ with $\omega_x$ decreases.
The behavior for the dissipative XXZ chain shows some interesting
differences. Taking $\ell\neq0$ increases the values of $I_d$,
but varying $\ell\neq$ does not induce appreciable variations on the
intrinsic dimensions. \textcolor{black}{We also considered the quantum state diffusion unraveling~\cite{Gisin_1992}. In this case, however, the Pareto assumption underlying the 2NN estimator appeared to break down. Since our analysis is numerical, a systematic exploration of all possible unravelings is beyond the scope of the present work; we therefore leave a more detailed investigation of alternative unravelings to future studies.} Nevertheless, in all scenarios analyzed here, the integrable point in parameter space is reliably identified by a minimum of $\overline{I_d}$.


\section{Conclusions}
\label{sec conclusions}

In this work we have investigated how the complexity of Linblad
dynamics is reflected in the structure of the corresponding quantum trajectories by determining their
intrinsic dimension. While the ensemble-averaged Lindblad dynamics
rapidly converges to a steady state which does not retain any memory
of the spectral structure of the Lindbladian, the trajectories are not
stationary and in general occupy lower dimensional subsets of the
Hilbert space. 
The dimensionality of these structures reflects the complexity of the
underlying Lindbladian, and shows signatures of properties that impose
constraints on the dynamics such as integrability or the decoupling
of the BBGKY hierarchy. It is not a priori expected that such constraints have an immediate effect on
the dynamics of QT because they are not associated with conservation
laws. Our work establishes that these constraints impose correlations
between trajectories which have the effect of reducing their complexity.

The integrable points of the dissipative quantum top are characterized by an intrinsic dimension of quantum trajectories equal to one, that increases in presence of perturbations that drive the model towards ergodic dynamics. The dependence of the intrinsic dimension on the model parameters mirrors that of spectral-statistic indicators of quantum chaos, including parameter regimes in which dissipative chaos emerges in the quantum model despite an underlying regular semiclassical limit, highlighting the intrinsically quantum nature of this effect.

For many-body models integrable trajectories remain
complex, but are consistently associated to local minima of the intrinsic dimension in the space of model parameters. Adding generic
perturbations in an ergodic regime does not significantly change the intrinsic dimension.
A qualitatively similar picture is obtained for dissipative models with a
decoupling point of their {BBGKY hierarchy}, where the intrinsic dimension is again observed to exhibit a local
minimum, signaling a drop in the complexity of
the underlying trajectories.

Altogether, our findings establish the intrinsic dimension as a
versatile data-driven probe of complexity in open quantum systems,
capable of distinguishing dynamical regimes beyond what is accessible
from ensemble-averaged observables alone, while also offering a bridge
towards notions of classical chaos. \textcolor{black}{At the current state, it should be viewed as a probe of dynamical compression, rather than as a standalone classifier of its microscopic origin. Whether finer geometric features of the dataset can further distinguish the microscopic origin of different constraints is an interesting question that we leave for future work.}

\section{Acknowledgements}

\textcolor{black}{We are grateful to Alessandro Laio for his helpful comments.} 
This work has been supported by the ERC under grant agreement
n.101053159  RAVE (RF, ET), by the PNRR MUR project PE0000023-NQSTI
(RF, MC), by the EPSRC under grant EP/X030881/1 (FHLE), and by the
PRIN 2022 (2022R35ZBF) - PE2 - ``ManyQLowD'' (MC). FHLE and RF thank
the Institut Henri Poincaré (UAR 839 CNRS-Sorbonne Université) and the
LabEx CARMIN (ANR10-LABX-59-01) for their support. 
E.T. was funded by the Swiss National Science Foundation (SNSF) under Grant No. TMPFP2234754. Moreover, E.T. acknowledges CINECA (Consorzio Interuniversitario per il Calcolo Automatico) award, under the ISCRA initiative and Leonardo early access program, for the availability of high-performance computing resources and support.

\appendix

\begin{figure*}[t!]
\centering

\includegraphics[width=0.95\textwidth]{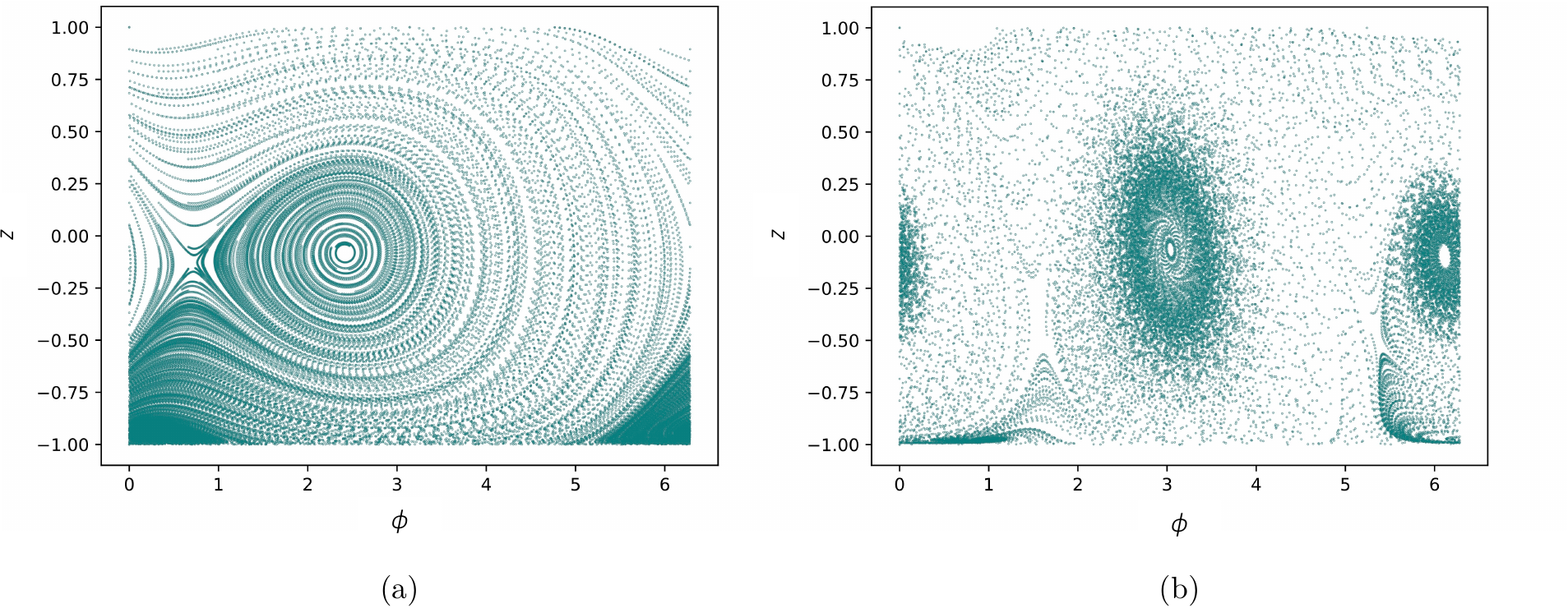}

\caption{Phase space orbits of the classical limit of the quantum top in Eq.~\eqref{eq: semiclassical system} for (a) $\omega_x=3, \,k=0$; (b) $\omega_x=2,\,k=0.2$. The flow is constrained on the sphere with constant $|\mathbf{m}|^2$, and the canonical variables are the standard cylindrical coordinates $z\equiv m_z,\,\cos\phi \equiv m_y/m_x$. The classical dynamics is regular as long as $k=0$, even when in presence of $\omega_x\neq0$ as in (a), complying with the Poincaré-Bendixson theorem. Introducing kicks results in the typical chaotic phase space shown in (b). } 

\label{fig: orbits}
\end{figure*}

\section{Semiclassical limit and the GHS conjecture}
\label{app: semiclassical limit}
%
The quantum top is semiclassical when the spin $S$ is large. In particular, it becomes equivalent to a classical continuous angular momentum, defined by the reduced magnetization operators $\hat m_\alpha = \hat{S}_\alpha/S$. In this limit the spectrum of each component becomes dense, and the angular momentum algebra turns into $[\hat m_\alpha,\hat m_\beta] = i \epsilon_{\alpha \beta \gamma}  \hat m_\gamma /S$, where $1/S$ becomes an effective Planck constant that vanishes for $S\to\infty$. The Lindblad Eq.~\eqref{eq: lindblad quantum top} implies the adjoint equation
\begin{equation*}
    \dv{}{t}\langle \hat{m}_\alpha \rangle = i \langle [\hat{H},\hat{m}_\alpha] \rangle + \frac{\gamma}{S} \langle \hat{S}_+ \hat{m}_\alpha \hat{S}_- -\frac12 \{\hat{S}_+\hat{S}_-,\hat{m}_\alpha\}\rangle\,.
\end{equation*}
Denoting $m_\alpha\equiv \langle \hat{m}_\alpha \rangle$, it is a
matter of algebra to show that the system, up to order $1/S$ corrections that arise from
$\langle \hat{m}_\alpha \hat{m}_\beta \rangle = m_\alpha m_\beta +
O(1/S)$, is equivalent to
\begin{subequations}
\label{eq: semiclassical system}
    \begin{align}
        & \dot m_x = \omega_z m_y -2g \,m_ym_z  + 2\gamma\, m_x m_z\\
        & \qquad \qquad \qquad + 2k\,\sum_n \delta(t-n\tau)\,m_ym_z \notag \\
        & \dot m_y = \omega_z m_x +2g\,m_xm_z -\omega_xm_z+2\gamma\,m_ym_z\\[+5pt]  
        & \dot m_z = \omega_xm_y -2\gamma(m_x^2 + m_y^2) \\
        & \qquad \qquad \qquad -2k\sum_n \delta(t-n\tau)\,m_xm_y
      \notag\ .
    \end{align}
\end{subequations}

A direct check shows that these equations always conserve the norm of
the classical spin so that the dynamics is constrained on the 2-sphere
at a fixed $|\vec{m}|^2 = m_x^2 + m_y^2 + m_z^2$.   
According to the Poincaré-Bendixson theorem, a classic result in the
theory of classical non-linear dynamics, in a two-dimensional
autonomous flow chaos is topologically forbidden
\cite{Wiggins2003}. The dynamics therefore must be regular when $k=0$
and the only possible attractors are fixed points and limit
cycles. Both are known to exist for $k=0,\,\omega_x=0$, where the
presence of periodic orbits is related to a boundary time crystal
phase \cite{Iemini2018}. When kicks are introduced the dynamics 
becomes explicitly time-dependent and chaotic regions appear, as can
be seen in Fig.~\ref{fig: orbits}.

\begin{figure*}[!]
\centering

    \includegraphics[width=0.95\textwidth]{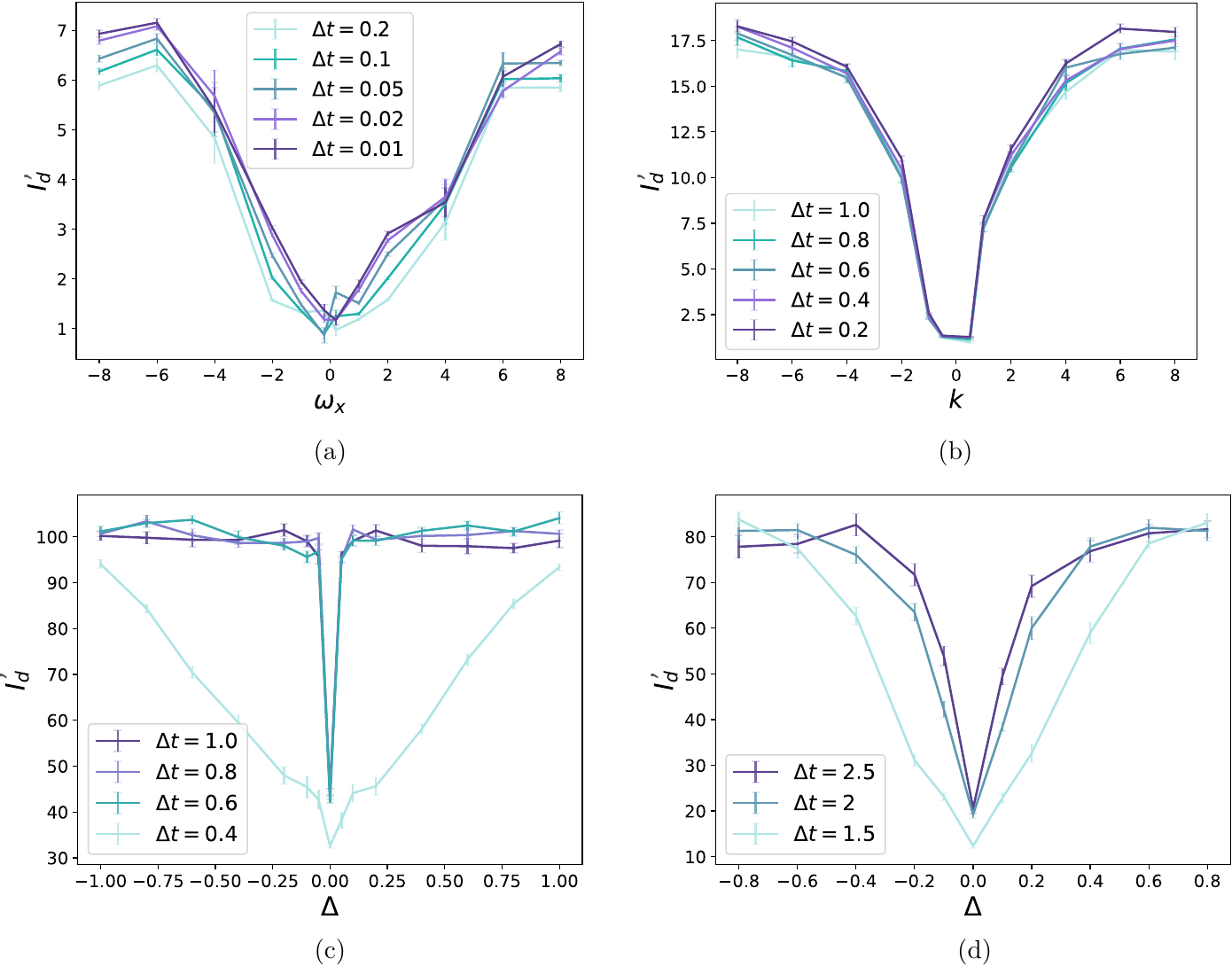}
    
\caption{Intrinsic dimension along individual trajectories. (a) and (b)
  Dissipative quantum top with  $\omega_x=1,\,g=5,\,\gamma=2$ and
  total spin $S=30$;
  (c) Model(A) with   $J_1=1,\,\gamma_0=1,\,J_2=0$ and length $L=8$;
  (d) Model (B) with $\gamma_1=1$. 
We calculate multiple estimations $I_d^{(i,\Delta
  t)}=I_d[\mathcal{D}_{i,\Delta t}]$ for each trajectory  $i=1,...,N$
using different values of $\Delta t$. The errorbars displayed
correspond to one standard deviation of the mean of the sample
$\{I_d^{(i,\Delta t)}\}_{i=1,...,N}$. The mean values of $I_d$ show
small variations with $\Delta t$, as they tend to increase when the
data set is observed at a finer scale able to capture “small”
additional dimensions. 
} 

\label{fig: Idst}
\end{figure*}

\section{Intrinsic dimension of individual quantum trajectories}
\label{app: Id single traj}

As we noted in Sec.~\ref{ssec: data sets} the intrinsic dimension of
QT can be calculated in at least two meaningful ways. Throughout
the main text of the paper we focused on the intrinsic dimension as a
function of time, i.e. the $I_d$ of the set of Hilbert space vectors
$|\psi_t\rangle_i\,,\,\,i=1,\dots,N$ reached by evolving $N$
independent QT from an initial state $|\psi_0\rangle$ up to the same
time $t$. Another natural approach would be to focus on individual
QT. Here the expectation is that, after an initial transient, QT
get confined to a manifold with dimension
$I_d<D=\dim_{\mathbb{R}}\mathcal{H}$. As a result the 
states $\{|\psi_0\rangle,\,|\psi_i({dt})\rangle,
\,|\psi_i({2dt})\rangle,\,\dots \,|\psi_i(t)\rangle\}$ of the $i^{\rm
  th}$ QT for sufficiently large $t$ will provide a discrete
approximation of the attracting manifold. The time step $\d t$ used to
simulate a QT is a microscopic quantity, and it can be convenient to
subsample the trajectory by considering only the state every $\Delta t
=k \cdot \d t$ for some integer $k\le1$. We calculate the intrinsic
dimension of the data set $\mathcal{D}_{i,\Delta t}=\{|\psi_0\rangle,\,|\psi_i({\Delta
  t})\rangle, \,|\psi_i({2 \Delta t})\rangle,\,\dots
\,|\psi_i(t)\rangle\}$ using the {\tt 2-NN} method, and the results
are then averaged over $N$ trajectories
{\color{black}\begin{equation}
  I'_d(\Delta t)=\frac{1}{N}\sum_{i=1}^N  I_d^{\rm 2-NN}[\mathcal{D}_{i,\Delta t}]\,.
\end{equation}
}
In this setting the ratio $k$ between
the time step $\Delta t$ used to construct the data set and the microscopic time
step $\d t$ of the simulation sets the scale on which the intrinsic
dimension is calculated. If $\Delta t$ is sufficiently small 
the one-dimensional structure of trajectories is expected to
emerge. Here we are interested in what happens when the total 
evolution time is long enough, so that the trajectory fill a
submanifold of the Hilbert space, and $\Delta t$ is large enough to
lose memory of the microscopic 1d structure.

\begin{figure*}[ht!]
\centering

    \includegraphics[width=0.9\textwidth]{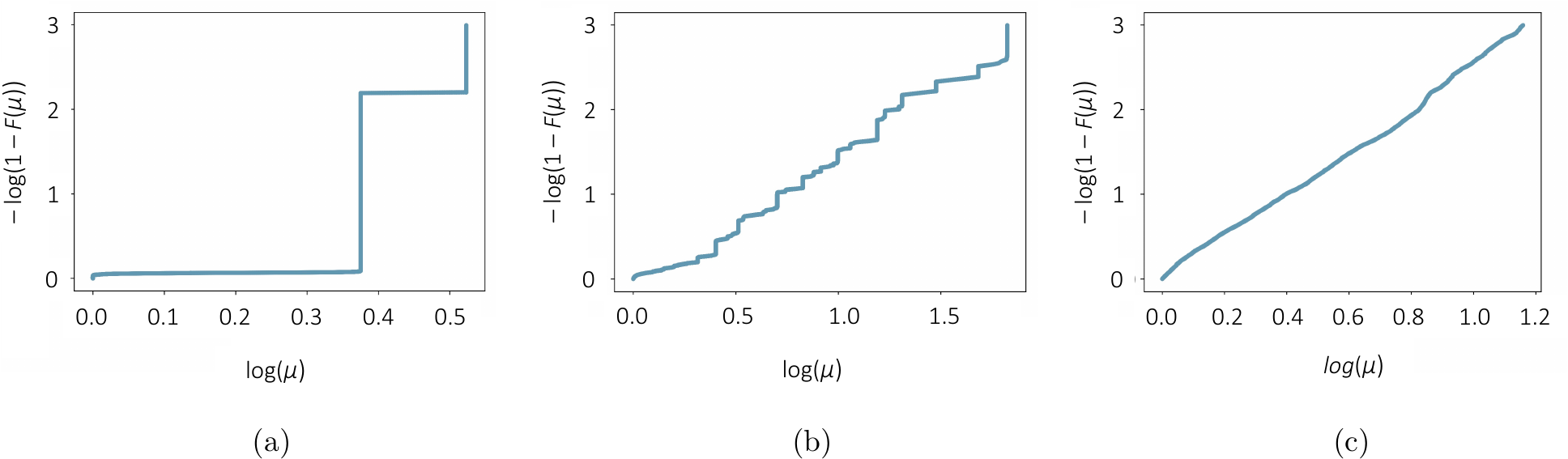}

\caption{Empirical cumulative distributions of the nnn/nn ratios for the dissipative quantum top with $\omega_z=1,\,g=5,\,k=0$, starting from (a) with $\omega_x=0$ to increasingly chaotic models (b) and (c), with $\omega_x=0.3$ and $\omega_x=1$. The axis are $\log\mu,\,-\log{(1-\hat{F}( \mu ))}$ so that for a Pareto distribution the graph is supposed to be linear. The distribution in (a) is very different from the ideal law and $I_d$ cannot be reliably estimated, but increasing $\omega_x$ a Pareto law is approached and even when the distribution is not linear yet, like for $\omega_x=0.3$, an effective $I_d$ can be estimated for most trajectories.} 

\label{fig: pareto}
\end{figure*}

In panels (a) and (b) of Fig~\ref{fig: Idst} we show the intrinsic
dimension along single trajectories for the quantum top as functions
of $\omega_x$ and $k$ obtained by averaging over $N=50$
trajectories starting from a fixed random state.
Fig~\ref{fig: Idst} (c) and (d) show the analogous results for
the dissipative XXZ chain.
Using the single-trajectory method we were unable to estimate $I_d$
at the integrable point on the quantum top $\omega_x=0,\,k=0$
because the hypothesis of a Pareto distribution for the ratios $\mu_i
= r^{nnn}_i/r^{nn}_i$ breaks down (cf. Sec.~\ref{ssec: intrinsic
  dimension}). This is a consequence of 
the great regularity of long-time integrable trajectories, for which
it is not correct to assume that nearest neighbors are extracted from
a Poisson process.  When a chaotic coupling is switched on, the local
structure of the time series changes and we observe the emergence of
complex higher-dimensional attractors in the Hilbert space with
$1<I_d\ll D$. Adding even a tiny amount of chaos is sufficient to
remove the extreme regularity of the integrable point and it becomes
possible to estimate the intrinsic dimension, as is shown in
Fig.~\ref{fig: pareto}.

This issue does not arise for the dissipative
XXZ chains, which instead display a greater sensitivity to the choice
of $\Delta t$. We observe that small values $\Delta t\sim10^{-2}$ do not result in
a Pareto distribution. For larger values $\Delta t\agt 10^{-1}$ 
we observe the expected scale dependence of the intrinsic dimension,
which arises due to the fact that by decreasing $\Delta t$ we probe a smaller scale and the values grow, signaling
a departure from the microscopic 1d structure.
The observed values for the intrinsic dimension and its functional dependencies
on $\omega_x$ and $k$ for the quantum top, and $\Delta$ for the
dissipative XXZ chain broadly comparable with the late-time averages
$\overline{I_d}$ shown in Figs.~\ref{fig: Idft   top}, \ref{fig: Idft
  XXZ},  confirming that the two methods capture the same submanifold.
Small deviations in the estimated values of the intrinsic dimension
between the two approaches are to be expected because of the scale
dependence, which is difficult to control quantitatively. A more
detailed discussion on the scale dependence is included in
Appendix~\ref{app: Id scaling}, focusing on the time-averaged
intrinsic dimension. 
%

\section{Scale dependence of the intrinsic dimension}
\label{app: Id scaling}

The intrinsic dimension of an unknown data manifold one aims
to learn depends on the scale at which the manifold is probed,
\emph{cf.} our discussion at the end of Section~\ref{ssec: intrinsic
  dimension}. For a nearest-neighbor based algorithm like the {\tt 2-NN}
method used in this paper, this typically results in a dependence on
the number of data points used to estimate $I_d$. Under the assumption
that additional data points will lie on the same submanifold,
generating more samples is expected to probe the data manifold on a
finer scale. Focusing on the intrinsic dimension at fixed times and on
the data set in Eq.~\eqref{eq: data set}, we explore this dependence
by increasing the number $N$ of QT  used to compute the
long-time average $\overline{I_d}$. To keep track of the scale we compute the
typical nearest neighbor distances $\overline{r}=1/2\,\cdot\,\overline{r^{nn}+r^{nnn}}$
in each estimation of $I_d(t)$.

\begin{figure*}[ht]
\centering
\includegraphics[width=0.95\textwidth]{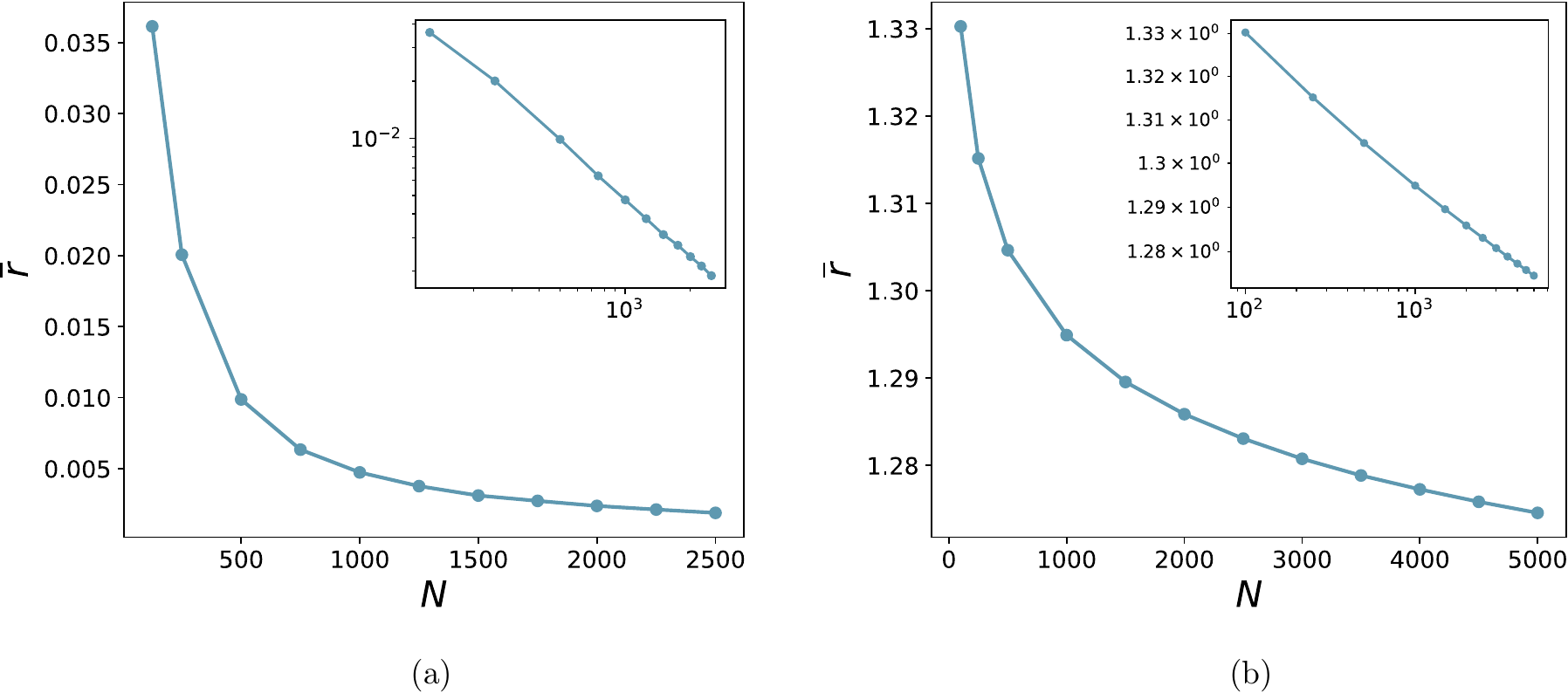}
\caption{Typical nearest-neighbor distance $\overline{r}$ as a function of the number of QT: (a)Quantum top with $S=30$ and $\omega_x=6.0$; (b) Model (A) with $L=8$ and $\Delta=1.0$. The inserts display the same data in log-log scale, highlighting the algebraic decay.}
\label{fig: r vs N}
\end{figure*}
In Fig.~\ref{fig: r vs N} we plot the
long-time average of $\overline{r}$, when the QT have settled and $I_d(t)$ has
relaxed. This results in a decreasing power-law scaling with $N$, compatible
with the expectation of a fixed submanifold that is being probed at a
finer scale.
\begin{figure*}[ht]
\centering

    \includegraphics[width=0.95\textwidth]{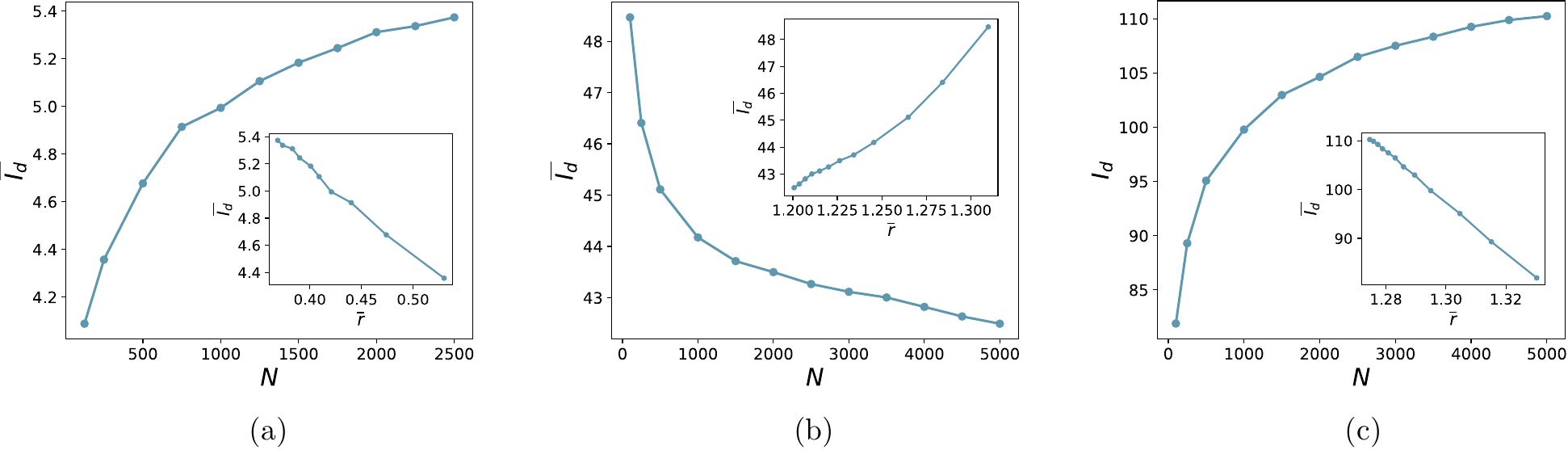}

\caption{Scale dependence of the intrinsic dimension, shown as functions of the number of QT and of the typical distances (inserts). (a) Chaotic quantum top with $S=30,\,\omega_x=6.0$. (b) XXZ chain subject to dephasing dephasing (Model (A)) with $L=8$ at its integrable point $\Delta=0$ and (c)  in its ergodic regime at $\Delta=1$.}
\label{fig: Id vs N}
\end{figure*}
The intrinsic dimension as a function of the scale is
shown in Fig~\ref{fig: Id vs N}. At the integrable point of the
quantum top $I_d\approx1$ regardless of the details, whereas chaotic
points exhibit an $N$-dependent $I_d$, as already noted in
Sec.~\ref{sec: quantum top}. The dimensionality increases with the
number of QT and displays an approximately linear dependence 
on the typical nearest-neighbor distances.

The behavior for the dissipative XXZ chain with finite anisotropy is
similar, while at the integrable point increasing the number of QT
results in a decreasing dimensionality, signaling an emerging
simplification of the data structure at a short scale.
We note that it would not be appropriate to extrapolate these
behaviors to a hypothetical $N\to\infty$ limit since the intrinsic
dimension is expected to eventually display a plateau (up to further 
deviations induced by noise) \cite{Facco2017, Denti2022}. In the range
of $N$ considered in our work there was no indication of such a plateau,
and significantly larger numbers of samples may be required to detect
one. \textcolor{black}{Nonetheless, the minima become deeper when more trajectories are considered both for the quantum top (integrable $I_d\approx1$, chaotic $I_d$ increasing with $N$) and for the XXZ chain with dephasing (integrable $I_d$ decreasing with $N$, chaotic $I_d$ increasing), signaling stability of our results. We note a different behavior for the minimum associated to the BBGKY decoupling, where Fig.~\ref{fig: Idft XXZ}~(d) shows complex and simplified points to all be decreasing when a finer scale is probed.}
A drawback of the approach we have used to analyze the scale
dependence of $I_d$ is that the statistical error due to the finite
number of samples increases at larger scales, and a more refined analysis could
be carried out with other algorithms, such as {\tt Gride}
\cite{Denti2022}.   

The main conclusion of our analysis of the scale-dependence of the
intrinsic dimension is that the reduction in the complexity of QT due
to constraints in the underlying Lindblad equation is observed over a
very wide range of scales.

\bibliography{bibliography.bib}
\end{document}